\documentclass[reprint,prx,aps,superscriptaddress]{revtex4-2}
\usepackage{times}
\usepackage{graphicx}
\usepackage{amsmath, braket, amsfonts}
\usepackage{amssymb}
\usepackage{natbib}
\usepackage{sidecap}
\usepackage{bm, color, ulem}
\usepackage{xcolor}
\usepackage{ragged2e}
\usepackage{float}
\usepackage{wrapfig,lipsum,booktabs}
\usepackage{siunitx}

\newcommand{\be}{\begin{equation}}
\newcommand{\ee}{\end{equation}}

\DeclareUnicodeCharacter{03BD}{{$\nu$}}
\DeclareUnicodeCharacter{2212}{{-}}

\makeatletter
\def\maketitle{
\@author@finish
\title@column\titleblock@produce
\suppressfloats[t]}
\makeatother

\begin{document}
\title{
Slow quasiparticle dynamics and anyonic statistics \\ in a fractional quantum Hall Fabry-P\'erot interferometer
}
\author{Noah L. Samuelson}
\thanks{These authors contributed equally to this work}
\affiliation{Department of Physics, University of California at Santa Barbara, Santa Barbara CA 93106, USA}
\author{Liam A. Cohen}
\thanks{These authors contributed equally to this work}
\affiliation{Department of Physics, University of California at Santa Barbara, Santa Barbara CA 93106, USA}
\author{Will Wang}
\affiliation{Department of Physics, University of California at Santa Barbara, Santa Barbara CA 93106, USA}
\author{Simon Blanch}
\affiliation{Department of Physics, University of California at Santa Barbara, Santa Barbara CA 93106, USA}
\author{Takashi Taniguchi}
\affiliation{International Center for Materials Nanoarchitectonics,
National Institute for Materials Science,  1-1 Namiki, Tsukuba 305-0044, Japan}
\author{Kenji Watanabe}
\affiliation{Research Center for Functional Materials,
National Institute for Materials Science, 1-1 Namiki, Tsukuba 305-0044, Japan}
\author{Michael P. Zaletel}
\affiliation{Department of Physics, University of California, Berkeley, California 94720, USA}
\affiliation{Material Science Division, Lawrence Berkeley National Laboratory, Berkeley, California 94720, USA}
\author{Andrea F. Young}
\email{andrea@physics.ucsb.edu}
\affiliation{Department of Physics, University of California at Santa Barbara, Santa Barbara CA 93106, USA}
\date{\today}
\begin{abstract}
Anyons are two dimensional particles with fractional exchange statistics that emerge as elementary excitations of fractional quantum Hall phases\cite{leinaas_theory_1977, wilczek_quantum_1982, arovas_fractional_1984, halperin_statistics_1984, goldin_particle_1980}. 
Experimentally, their exchange statistics can be measured in the edge-state Fabry-P\'erot interferometer \cite{kivelson_semiclassical_1990,chamon_two_1997,kim_aharanov-bohm_2006,halperin_theory_2011}.  In these devices, the presence of $N_{qp}$ localized anyons in the bulk contributes a phase $N_{qp}\theta_a$ to the interference pattern where $\theta_a$ is twice the exchange phase\cite{read_clarification_2024}.
Here we report the observation of large, hysteretic phase jumps in a monolayer graphene Fabry-P\'erot interferometer at $\nu=1/3$.  
When the filling factor is increased from $\nu<1/3$ towards the center of the plateau, we observe phase slips with magnitude $\Delta \theta \approx 2\pi/3$, consistent with the addition of individual quasiparticles to the interferometer bulk as observed in both GaAs and graphene\cite{nakamura_direct_2020, nakamura_fabry-perot_2023, werkmeister_anyon_2025}.  
In contrast to prior work, however, the phase slips occur as instantaneous jumps in the interference signal, indicative of quasiparticle equilibration times exceeding 20 minutes.  
We use this long timescale to investigate the effect of changes in interferometer area $A_I$ and $N_{QP}$ independently at fixed magnetic field, revealing a striking memory effect in the phase slip magnitude.  
In particular, as the $\nu=1/3$ plateau is approached from higher filling, we observed phase slips with $\Delta \theta$ significantly larger than $2\pi/3$ over the same range of gate voltage where quantized jumps are seen for increasing $\nu$. 
We discuss this asymmetry in terms of bulk-edge coupling of quasiparticles localized near the edge or in the bulk, and argue that this effect can be qualitatively reconciled with theoretical expectations for strongly interacting quasiparticles in the presence of weak disorder and strongly nonequilibrium charge dynamics.
Besides providing a replication of interferometric measurements sensitive to $\theta_a$ \cite{nakamura_direct_2020, werkmeister_anyon_2025}, our results highlight the key role played by charge dynamics on signatures of the anyon phase, and demonstrate that fractional quasiparticles can be indefinitely localized in nonequilibrium configurations.  
\end{abstract}

\maketitle

\section{Introduction}

\begin{figure}[ht!]
    \includegraphics[width = 90mm]{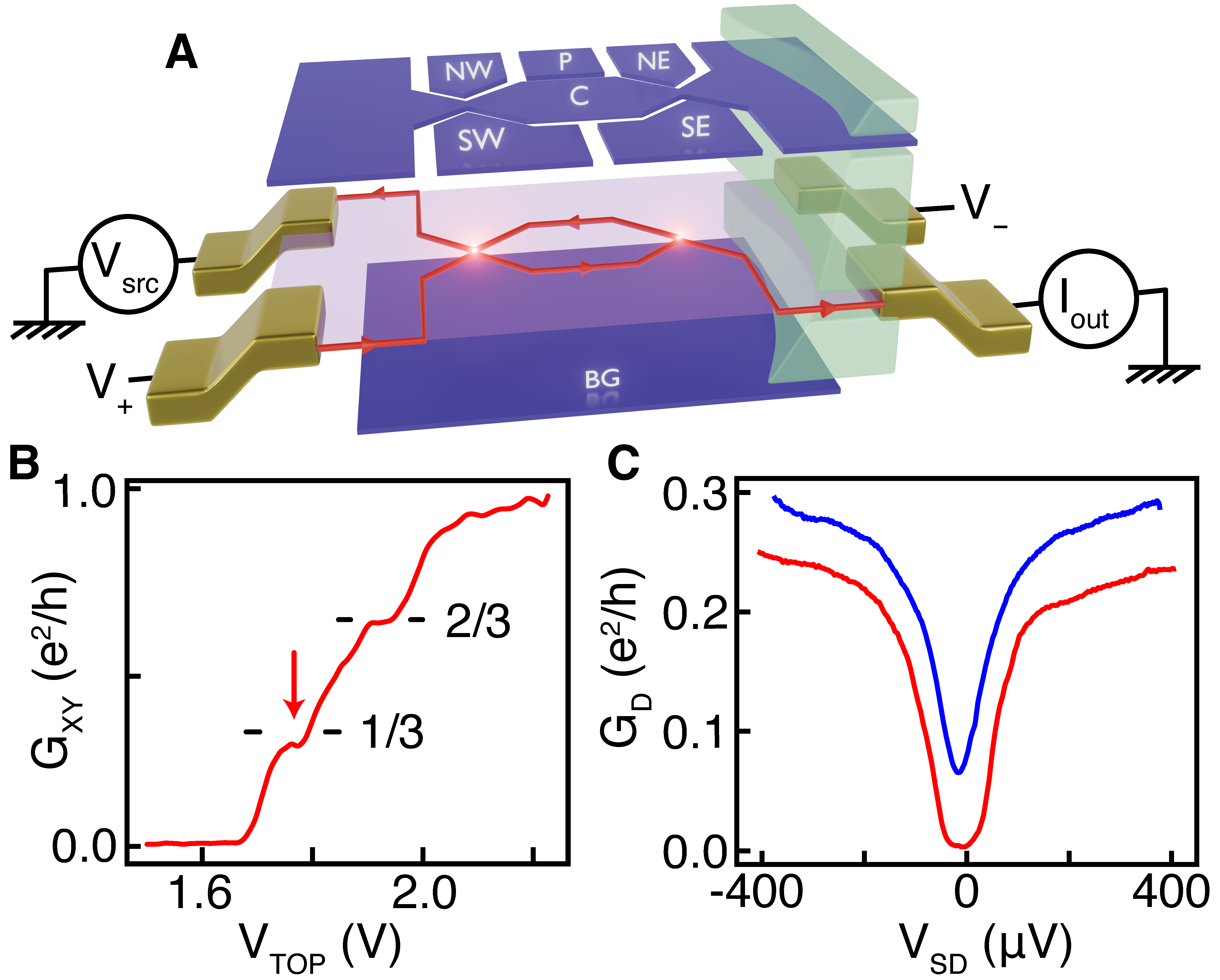}
    \caption{
\textbf{Device schematic and characterization in the $\nu = 1/3$ state}
\textbf{(A)} Schematic of the dual-graphite gated edge state Fabry-P\'erot interferometer.  
The gates defining the interferometer are labeled C, NW, SW, NE, SE and P.  Edge states are formed in the monolayer graphene around the C-gated region and enter the interferometer via two QPCs tuned by the NW/SW and NE/SE gates. There is also a global graphite back gate (BG).
\textbf{(B)} Hall conductance $G_{XY}$ measured on the west side of the device as a function of the voltage $V_{TOP}$ applied to all top gates together, with a fixed back gate voltage $V_{BG} = \SI{-1.5}{V}$ at $B=\SI{9}{T}$ and $T=\SI{18}{mK}$.  
\textbf{(C)} Conductance through each QPC vs. the DC source-drain bias $V_{SD}$ (applied as an added voltage on the source electrode), with the $V_{BG}=\SI{-2.0}{V}$ throughout.  The blue trace was taken with the C/P/NW/SW regions at filling 1/3 ($V_{C/P/NW/SW}=\SI{2.232}{V}$), and the NE/SE regions depleted ($V_{NE/SE}=\SI{-4.1}{V}$).  The red trace was taken with the NE/SE regions at filling 1/3 ($V_{C/P/NE/SE}=\SI{2.232}{V}$) and the NW/SW regions depleted ($V_{NW/SW}=\SI{0.5}{V}$). See Supplement for the QPC pinch-off curves as a function of $V_{NW/SW}$ and $V_{NE/SE}$. 
\label{fig:fig1}}
\end{figure}

When an Abelian anyon is brought along a closed trajectory encircling $N_{qp}$ localized anyons, its wavefunction accumulates a phase 
\begin{equation}
    \frac{\theta}{2\pi} = \frac{e^\ast}{e} \frac{A_I B}{\Phi_0} +  N_{qp} \frac{\theta_a}{2\pi},
\label{theta}
\end{equation} where $A_I$ is the area of the loop, $B$ is the applied magnetic field, $\theta_a$ is twice the exchange phase, and $e^\ast$ is the quasiparticle charge. Quantum Hall edge state Fabry-P\'erot interferometers exploit the contrast between localized anyons in the bulk and propagating anyonic quasiparticles along the chiral edge modes to directly observe this phase\cite{kivelson_semiclassical_1990,chamon_two_1997,kim_aharanov-bohm_2006,halperin_theory_2011}. 
In a Fabry-P\'erot interferometer, delocalized quasiparticles enter the cavity via a quantum point contact (QPC) and propagate along the edge to a second QPC; they can then exit the cavity immediately or complete an integer number of additional circuits before exiting. 
Trajectories differing by the number of circuits give an interference contribution to the conductance,  $\delta G$, that is periodic in $\theta$ and can be measured as a function of $B$, $A_I$, or $N_{qp}$.  
The clearest signature of anyonic statistics is expected if $N_{qp}$ changes discretely while keeping $A_I$ and $B$ fixed; the resulting jump in $\theta$ then gives $\theta_a$ directly. 

\begin{figure}[hb]
    \includegraphics[width = 90mm]{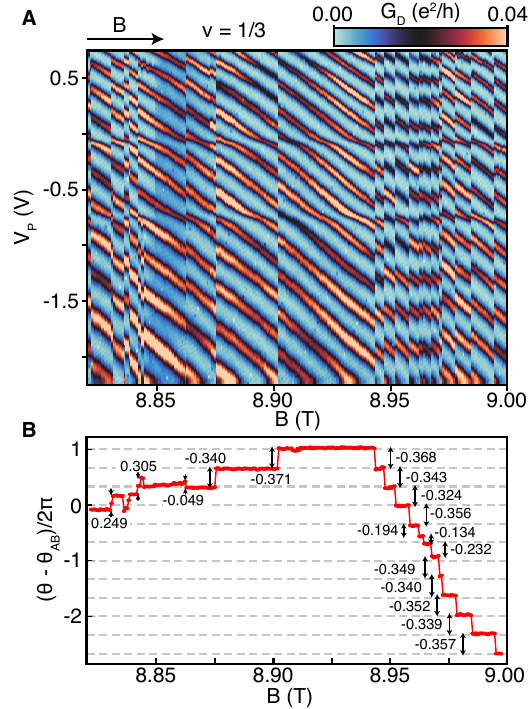}
    \caption{
\textbf{Fabry-P\'erot interference at $\nu = 1/3$}
\textbf{(A)} $G_D$ measured across the interferometer at $\nu = 1/3$, with both QPCs partially transmitting. The magnetic field is swept from low to high.
\textbf{(B)}  The phase extracted from the Fourier transform of the data in panel D. A continuous Aharonov-Bohm phase is subtracted, and we adopt the convention $\Delta \theta \in (-\pi,\pi)$.  
\label{fig:fig2}}
\end{figure}

Fabry-P\'erot interferometers have been investigated in GaAs heterostructures for nearly two decades\cite{camino_realization_2005,
camino_e/3_2007,zhang_distinct_2009,
mcclure_edge-state_2009, ofek_role_2010, choi_robust_2015,sivan_observation_2016, nakamura_aharonovbohm_2019,nakamura_direct_2020, nakamura_fabry-perot_2023, willett_interference_2023}. 
These experiments have revealed that Coulomb interactions complicate the naïve interpretation that a jump in $\theta$ entirely arises from the anyonic exchange phase\cite{halperin_theory_2011}.  Specifically, as charge enters the bulk of the interferometer, Coulomb repulsion may cause a change in $A_I$, leading to an observable phase shift even for fermionic quasiparticles. 
If the ``bulk-edge coupling'' is large, this change in the Aharonov-Bohm phase can completely obscure the contribution of $\theta_a$. 
Recently, a breakthrough in the design of GaAs heterostructures led to the observation of phase shifts that agree quantitatively with the expected $\theta_a = 2\pi/3$ in the $\nu$ = 1/3 state\cite{nakamura_aharonovbohm_2019,nakamura_direct_2020,nakamura_impact_2022, nakamura_fabry-perot_2023}.  Graphene heterostructures are a natural venue in which to extend these results owing to the large fractional quantum Hall energy gaps observed at both odd-\cite{dean_multicomponent_2011,polshyn_quantitative_2018,zeng_high-quality_2019,dean_fractional_2020} and even-denominator filling factors\cite{zibrov_tunable_2017, li_even_2017, zibrov_even-denominator_2018, hu_high-resolution_2025,assouline_energy_2024}.  Moreover, the nearby graphite gates in typical dual-gated geometries ensure a high degree of screening, suppressing bulk-edge coupling.  Indeed, measurements of Fabry-P\'erot interferometers in graphene have uniformly observed Aharonov-Bohm dominated interference in both integer and fractional filling factors \cite{deprez_tunable_2021, ronen_aharonov-bohm_2021, zhao_graphene-based_2022,fu_aharonovbohm_2023,kim_aharonovbohm_2024, werkmeister_anyon_2025}.

\section{Fabry-P\'erot Interference in $\nu=1/3$ }

In this work, we study a monolayer graphene gate-defined Fabry-P\'erot interferometer, shown schematically in Fig. \ref{fig:fig1}a. The interferometer is fabricated using anodic oxidation lithography to define the gate structure in a graphite layer.  Dry van der Waals assembly techniques are then used to form a device with six separately-gated regions\cite{cohen_nanoscale_2023}. In our device (Fig. \ref{fig:fig1}a), two pairs of gates (NE/SE and NW/SW) define QPCs, while a plunger gate (P) provides additional control over the interferometer area.  The center gate (C) and a global graphite bottom gate are used, together, to set the filling factor and, through fringe electric fields, adjust the transmission through the QPCs.  All data are measured in a dry dilution refrigerator with a base temperature of 17mK.  Transport data at $B=\SI{9}{T}$ as a function of a common voltage applied to all six top gates in Fig. \ref{fig:fig1}b show well-developed plateaus at filling factor 1/3 and 2/3.  
We operate our interferometer within the 1/3 plateau at the indicated point at $B=\SI{9}{T}$. 
We measure the transmission through the interferometer via the diagonal conductance, $G_D \equiv I_{\text{out}} / (V_+ - V_-)$ (see Fig.~\ref{fig:fig1}a).  
To confirm that our experiment is probing chiral edge modes of the 1/3 state, we measure the source-drain bias dependence of the two QPCs in the partial transmission regime individually.  As shown in Fig.~\ref{fig:fig1}c, both show strong suppression of $G_D$ at low bias, as expected for tunneling between chiral Luttinger liquids at the QPCs\cite{wen_edge_1991, kane_transport_1992,  fendley_exact_1994, kane_impurity_1995,  fendley_exact_1995, milliken_indications_1996, radu_quasi-particle_2008, cohen_universal_2023}. This behavior is neither expected nor observed in the integer quantum Hall regime (see Supplement for a comparable measurement in the $\nu = 2$ state, showing no zero-bias suppression).

Fig.~\ref{fig:fig2}a shows $G_D$ at $\nu=1/3$ with both QPCs set to partial pinch-off as a function of $V_P$ and $B$. 
The interference shows high-visibility oscillations with lines of constant phase having a negative slope in the $V_P-B$ plane, consistent with an Aharonov-Bohm dominated interference phase\cite{halperin_theory_2011}.
Following Eq.~\eqref{theta}, we estimate the effective interferometer area to be $A_I = 3 \Phi_0 / \Delta B = 0.69-0.83 \mathrm{\mu m^2}$ based on the  $\Delta B\approx 15-\SI{18}{mT}$ field period of the oscillations.  
This agrees with the nominal device area of $0.74-0.83 \mathrm{\mu m^2}$ of the patterned graphite gates (see supplement for a more detailed analysis of the magnetic field period).  Estimates of the edge state velocity (see supplementary information) give $v = 6.2\pm 0.2\times 10^4 \mathrm{m/s}$, comparable to prior estimates for integer quantum Hall edge states in graphene\cite{deprez_tunable_2021,ronen_aharonov-bohm_2021,fu_aharonovbohm_2023,werkmeister_strongly_2024}.  We also characterize the visibility of Aharanov-Bohm interference at $\nu = 1/3$ as a function of temperature where we extract a characteristic temperature scale $T_0 = \SI{87}{mK}$ (see supplement).

In addition to the continuously-tuned Aharonov-Bohm phase giving rise to the negative slope, the most striking feature of Fig. \ref{fig:fig2}a is the presence of `hard' phase slips, where the interference phase changes  instantaneously with respect to the 30-second measurement time of each individual $V_P$ trace, resulting in an apparent discontinuity between traces taken at subsequent magnetic fields.
We also observe two clear `soft' phase slips, located at $V_P = \SI{-0.1}{V}$ and $V_P = \SI{-0.8}{V}$, which are tuned continuously with $V_P$.
These phase slips are nearly horizontal in the $V_P  - B$ plane, indicating a strong degree of capacitive coupling to the plunger gate likely from a localized defect state near the interferometer boundary.
Defects of this type have been investigated in the integer quantum Hall regime in a previous work \cite{samuelson_hard_2025}. 
Here we focus on the `hard' phase slips which constitute the majority of the observed events.

To quantify the magnitude of the phase slips, we compute the Fourier transform of $G_D$ with respect to $V_P$ (over the range $V_P\in [\SI{-2.25}{V}, \SI{-0.78}{V} ]$ in order to avoid the effect of the `soft' phase slips) for each value of $B$ and extract the phase of the largest-magnitude peak, which determines the oscillation phase $\theta$. 
Per Eq. \eqref{theta}, $\theta$ is expected to contain both a smoothly-varying Aharonov-Bohm contribution as well as discrete contributions proportional to $N_{qp}$.  The latter term includes the anyon double-exchange phase, as in \eqref{theta}, as well as the effect of bulk edge coupling, which induced $N_{QP}$-dependent shifts in $A_I$\cite{halperin_theory_2011}. 
To isolate the $N_{QP}$ dependent terms, we take a running trimmed mean of the line-by-line phase differences and subtract it from the measured phase (see Supp Mat). 

The residual phase, $\theta - \theta_{AB}$, is plotted in Fig.~\ref{fig:fig2}b.
As this only determines the phase difference modulo $2\pi$, we adopt the convention $\Delta \theta \in (-\pi,\pi)$. 
The magnitude of each phase slip is calculated from the difference in average value between intervals of stable phase. 
The statistical error in the phase slip measurements is small; repeated measurements within a single interval of stable phase show a standard deviation of $\sigma_\theta \approx .012 \times 2\pi$. A larger source of error arises from the fact that the oscillations are not perfectly periodic in $V_P$, generating different values of $\theta$ for different components of the Fourier transform. 
We estimate that this error may be as high as $\pm2\pi\times .04$ (see Supplement).
Most of the marked phase slips are consistent, within this uncertainty, with $\Delta\theta = \pm 2\pi/3$.  

\begin{figure}[ht!]
    \includegraphics[width = 90mm]{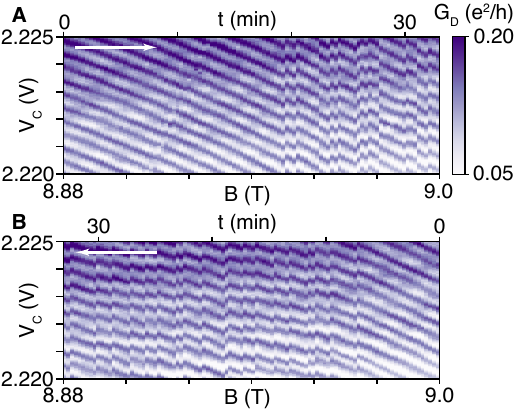}
    \caption{
\textbf{Magnetic field hysteresis.}
\textbf{(A)} $G_D$ vs. $V_C$ as the magnetic field is swept from 8.88T to 9T. The latter half of the measurement shows many sudden phase jumps, beginning at $B = 8.95T$ and continuing until the end of the sweep.
\textbf{(B)} $G_D$ vs. $V_C$ as the magnetic field is swept from 9T to 8.88T. The hard phase slips differ in exact location as compared to panel a; in particular, they do not begin until $B < 8.97T$ and then continuing until the end of the sweep at $B=8.88T$.
\label{fig:fig3}}
\end{figure}

In GaAs, observed phase slips are always soft: they evolve continuously with both $B$ and gate-voltage\cite{nakamura_direct_2020}.  This is consistent with an equilibrium picture in which quasiparticles move between the interior and exterior of the interferometer loop on time scales much faster than the integration time of the measurement.  Our observation in Fig. \ref{fig:fig2}a of hard slips, which appear instantaneous on measurement time scales, is not consistent with this picture, instead implying that quasiparticle dynamics are irreversible.  This is observed directly in Fig. \ref{fig:fig3}a-b, where we contrast interference plots taken as a function of increasing and decreasing $B$.  
In these plots, $V_P$ is fixed while $V_C$ is swept rapidly to vary the interferometer area and produce an interference pattern, while $B$ is ramped from minimum to maximum over 33 minutes.  
In Fig. \ref{fig:fig3}a, a series of discrete slips are observed; However, phase slips do not occur until the magnetic field has already been increased by $\SI{70}{mT}$, nearly 20 minutes after the beginning of the measurement. 

Reversing the direction of the $B$ sweep (Fig. \ref{fig:fig3}b) reveals a hysteretic behavior where phase slips start only after the field is \textit{lowered} sufficiently.
Evidently, the location of a given phase slip in our experiment is not an equilibrium property, but depends on the history of the system.  This suggests that the anyon occupation $N_{qp}$ must be considerably out of equilibrium with the sample edge. 
Notably, the Aharonov-Bohm phase at the beginning of the measurements in both Fig.~\ref{fig:fig3}a and b evolves through several periods as a function of $V_C$. This implies that the area changes by multiple flux quanta through the interference loop at fixed anyon charge and approximately fixed size.  This is not compatible with an equilibrium picture: $V_C$ is varied over a range several times larger than the expected charge gap, so it is energetically favorable to add quasiparticles to the interferometer.
\begin{figure}[ht!]
    \includegraphics[width = 70mm]{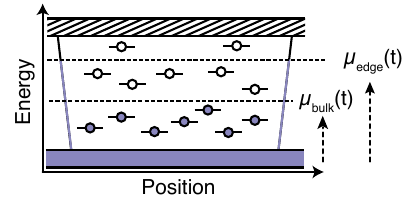}
    \caption{
\textbf{Irreversible charging dynamics.}
Localized states within the fractional quantum Hall energy gap have slow tunneling rates to the gapless edge; as a result, the  chemical potential of the electrically isolated bulk ($\mu_{bulk}$)  lags the chemical potential of the edge ($\mu_{edge}$) during a ramp of the magnetic field or gate voltage.  In this schematic, purple indicates filled states.  
\label{fig:fig4}}
\end{figure}

Hysteretic behavior of this type can be understood as a consequence of slow charge dynamics. As illustrated in Fig. \ref{fig:fig4},  slow charging leads to a discrepancy between the equilibrium value of the chemical potential, $\mu_{eq}$, which is set by the gate voltages and magnetic field, and the actual chemical potential of the graphene layer $\mu(t)$.
In a previous experiment on a device of nearly identical construction, we investigated this phenomenon in the integer quantum Hall regime. There, the observed phase-slips (which in that case arise entirely from bulk-edge coupling) originated from a well understood equilibrium charging picture, but became sudden-in-time and hysteretic as we continuously adjusted the dynamical barrier to charging the bulk with density and magnetic field\cite{samuelson_hard_2025}.
Thus, while it is remarkable to see this effect at the single fractional quasiparticle level, its occurrence is not completely unexpected.
Long charging times are a characteristic of quantum Hall systems\cite{goodall_capacitance_1985,eisenstein_negative_1992}, particularly in similarly fabricated graphene devices which boast low-disorder and larger energy gaps for quantum Hall states compared to competing platforms, such as GaAs quantum wells \cite{yang_experimental_2021}. 

The hysteretic, sudden-phase slip behavior is distinct from the 3-state random telegraph noise observed in a parallel work studying a similar dual-gated graphene interferometer \cite{werkmeister_anyon_2025}. There, phase slips occur indefinitely as a function of time even when the magnetic field and gate voltage are fixed, indicating fluctuations in the number of anyons while the system remains close to equilibrium. In our experiment, no phase slips occur when the gate voltage and magnetic field are held fixed over timescales of hours, or even when they are swept over a sufficiently small range (see Supplement). Instead, the system must be perturbed significantly by a change in either a gate voltage or magnetic field in order to observe repeatable phase slips.

\begin{figure*}
    \includegraphics[width = 180mm]{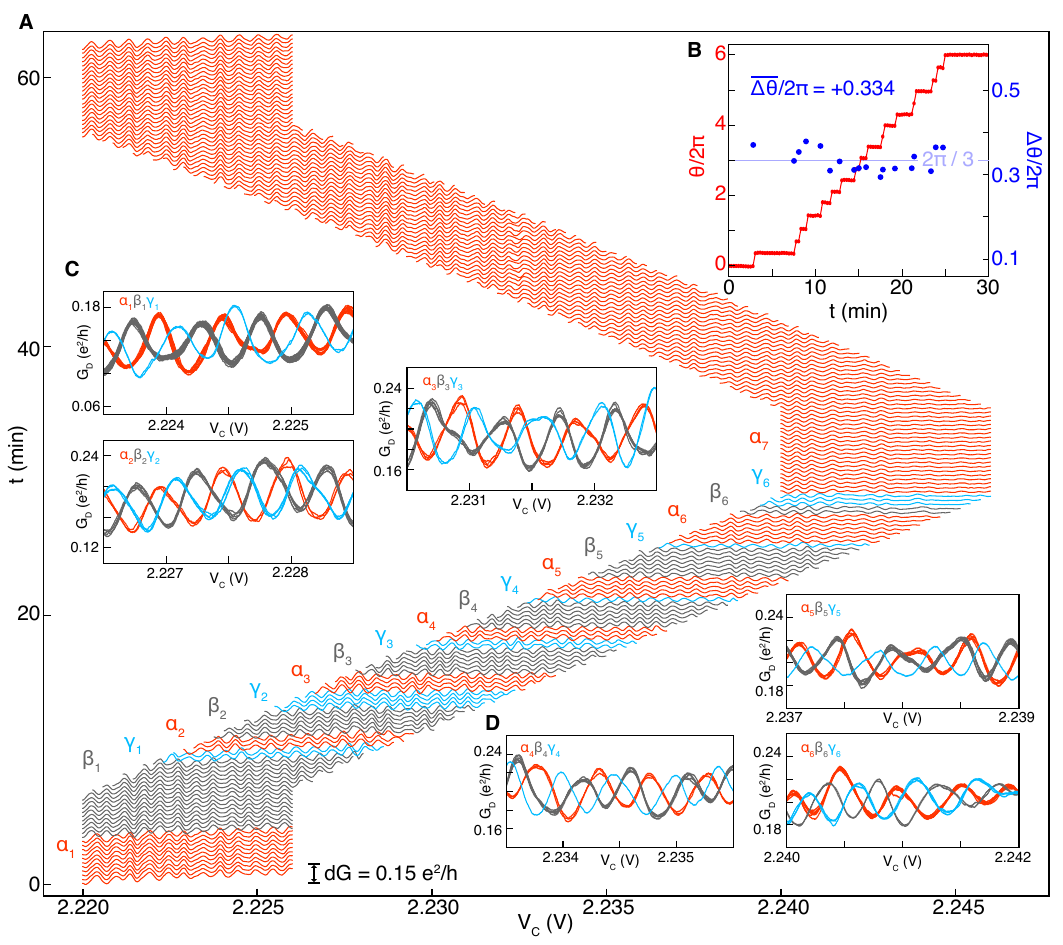}
    \caption{
\textbf{
Gate hysteretic phase slips at constant field.}
\textbf{(A)} Repeated line traces of the conductance $G_D$ plotted as a function of $V_C$. Traces are vertically offset by an amount proportional to the time between the start of each trace. The window over which $V_C$ is swept gradually increases over time, favoring an increase in the number of quasiparticles. Interpreting each phase slip as the addition of a single anyon, the traces are grouped into distinct classes $\alpha$, $\beta$, and $\gamma$ corresponding to the respective values of $N_{qp} \mod 3$. No phase slips are seen as the gate voltage is swept back over the range in the reverse direction.
\textbf{(B)} The phases extracted from the Fourier transform of each trace in panel A are plotted in red. Each phase jump is assumed to lie in the interval $(-\pi, \pi)$. Blue points correspond to the magnitude of each sudden jump in $\theta$.
\textbf{(C-D)} Collections of line traces from each set of three adjacent classes $\alpha_i$, $\beta_i$, $\gamma_i$ plotted over the region where the data overlap in $V_C$. The stability of the phase is apparent given the overlap between traces within a given class, while the classes are offset from each other by a $\sim 2\pi/3$ phase shift.
\label{fig:fig5}}
\end{figure*} 

The long charging time allows us to measure the phase slips systematically at \textit{fixed} magnetic field. This is accomplished by modulating a single gate voltage faster than the the charging time over a small range that nevertheless allows  us to measure the interference phase, and then slowly varying the center of that gate voltage range to increment $N_{qp}$.
Fig. \ref{fig:fig5}a shows data taken in this way at $B=9T$ as a function of $V_C$.  In this measurement, $V_C$ is swept over a range spanning roughly 10 oscillations in approximately 30 seconds. The range is adjusted from trace to trace, so that the average value of $V_{C}$ increments slowly over about one hour.  Most successive traces show identical oscillatory patterns, but a pattern of abrupt phase slips is again evident.  We label adjacent traces with the same phase with the same color, highlighting the stability of the phase over multiple scans.  
Notably, hysteresis similar to that observed as a function of magnetic field is again evident: a series of phase slips are visible for increasing $V_C$, but the phase remains stable as $V_C$ is decreased back over the same range.

Fig. \ref{fig:fig5}B shows the  phase $\theta$ of each trace extracted from the discrete Fourier transform. Notably, no background subtraction is necessary at constant $B$. We observe 18 phase slips over this range with a mean value of $2\pi\times 0.334$ and a standard deviation of $2\pi\times 0.038$. 
We interpret each of these slips as the entry of a single $e/3$ quasiparticle into the interferometer (or the neutralization of a single $-e/3$ quasi-hole).  We correspondingly collect traces into groups where the phase is the same, and label these by a Greek letter $\{\alpha, \beta, \gamma\}$ corresponding to $N_{qp} \mod 3$. 
A numerical subscript distinguishes traces with the same $N_{qp} \mod 3$ that are separated by 3 phase slips; assuming each phase slip corresponds to entry of a single fractionally-charged quasiparticle, these traces correspond to charge configurations differing by an integer number of whole electron charges in the interferometer. 
Fig. \ref{fig:fig5}C-D compares traces $\{\alpha_i,\beta_i,\gamma_i\}$ for $i=\{1,...,6\}$.  Each comparison shows a clear `triple-helix' pattern arising from the $2\pi/3$ relative phase shift between each set of curves.  

\section{Memory effect and Anomalous Phase Slip Magnitudes}

\begin{figure*}
    \includegraphics[width = 155mm]{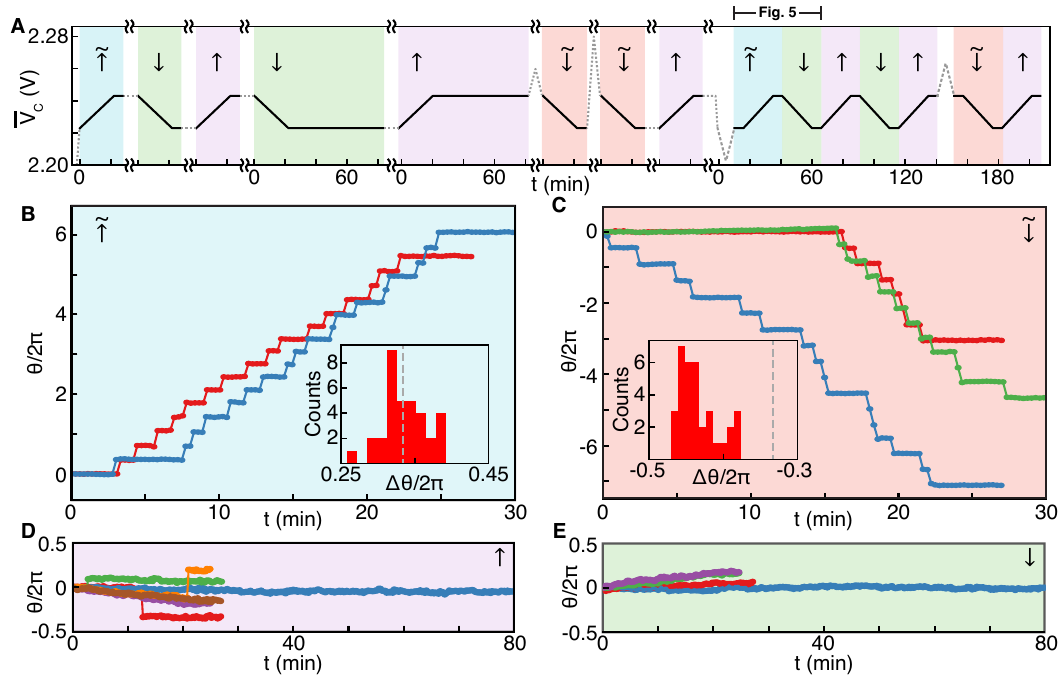}
    \caption{
\textbf{Dependence of phase slip phenomenology on gate history.}
\textbf{(A)}  Data is collected as the center gate voltage is ramped over a 6mV-wide window centered at the value of $\bar{V_C}$ in the illustrated trajectory.
\textbf{(B)} The phase extracted via FFT for sweeps of increasing $\bar{V_C}$ in which the measurement was preceded by an excursion to lower voltage outside the plateau ($\tilde \uparrow$),  
\textbf{(C)} decreasing $\bar{V_C}$ in which the measurement was preceded by an excursion to higher voltage outside the plateau ($\tilde \downarrow$),
\textbf{(D)} increasing $\bar{V_C}$ in which the measurement was not preceded by an excursion to lower voltage ($\uparrow$), and 
\textbf{(E)} decreasing $\bar{V_C}$ in which the measurement was not preceded by an excursion to higher voltage ($\downarrow$). 
Insets in panels B and C show histograms of the values of the phase differences $\Delta \theta$ between plateaus of stable $\theta$, with the gray dashed lines indicating $ \Delta \theta = \pm 2\pi/3$. In panel B, $\Delta\theta / 2\pi = +0.336 \pm 0.03$ and in panel C, $\Delta\theta = -0.434 \pm 0.03$.
\label{fig:fig6}}
\end{figure*}

The large hysteresis observed in Fig. \ref{fig:fig5} raises broader questions about the role of memory effects in quasiparticle dynamics and phase-slip phenomenology.
To investigate this, we perform repeated measurements following the protocol of Fig. \ref{fig:fig5}, in which the interferometer phase is measured by rapid rastering of $V_C$ in a small range while the center of this window, denoted $\bar{V_C}$, is slowly changed within the transport plateau.  
Besides performing multiple repetitions of the same trajectories, we also intersperse these measurements with `excursions' to either higher or lower $\bar{V_C}$ outside the transport plateau. This measurement protocol is shown in Fig. \ref{fig:fig6}a.  During the intervals indicated by the solid black line, interferometeric phase data is collected continuously by rasterizing $V_C$ while $\bar{V}_C$ is adjusted or held fixed. 
During the regions denoted by dotted black lines,   $\bar{V}_C$ is either held constant or swept over the indicated range \emph{without} taking interferometry data (in these dashed regions $V_C$ is not rasterized).  While these excursions are not precisely timed, they are typically several minutes in duration. The magnetic field and all other gate voltages are held constant for the entire course of the experiment.  Repeated measurements of the interferometer phase in the same gate voltage regime ($\bar V_C\in(2.223,2.243)$) are color coded two criteria: (1) whether they are measured for rising or falling $\bar{V_C}$ (and thus average $\nu$) and  (2) whether they are directly preceded by an excursion to large or small $\bar{V_C}$ before the falling or rising trajectories, respectively. 
We denote the four scenarios by $\uparrow, \downarrow, \tilde{\uparrow}, \tilde{\downarrow}$, where the arrow indicates the sweep direction, and a ``$\sim$'' indicates that the sweep was directly preceded by an excursion outside the transport plateau. 
Phase slip data for each of the resulting four sweep types is shown in Figs. \ref{fig:fig6}b-e.  

A striking pattern is evident in the presence or absence of phase slips for a given measurement. 
Specifically, multiple phase slips are observed consistently upon entry to the plateau from larger or smaller $\bar{V_C}$ (types $\tilde{\uparrow}, \tilde{\downarrow}$), but none are observed when the gate voltage direction is reversed within the plateau (types $\uparrow, \downarrow$). Consequently, multiple phase slips are observed for increasing $\bar V_C$ only on the first entry of the plateau from below ($\tilde{\uparrow}$, Fig. \ref{fig:fig6}b), but are observed for decreasing $\bar V_C$ only for the first entry of the plateau from above ($\tilde{\downarrow}$, Fig. \ref{fig:fig6}c). 
With the exception of the occasional events observed in Fig. \ref{fig:fig6}d,  slips are never observed upon repeated reversal within the plateau for either sweep direction  ($\uparrow, \downarrow$)after the first sweep until the voltage is again swept well outside the central range.  

The presence of hysteresis makes it clear that  different microscopic charge configurations can be accessed at the same values of $V_C$ depending on the history of the system. This is most obvious in the irreversible ratcheting of $N_{qp}$, which is consistent with a mobility gap near the plateau center where charge equilibration times become far larger than experimentally accessible.  Less obvious is the fact that rising and falling-$\bar V_C$ traces may access radically different charge configurations.  The large excursions in $\bar{V_C}$ preceding a rising or falling trace will tend to heavily dope the interferometer; $\tilde\uparrow$ sweeps thus begin from a state with many quasiholes while $\tilde \downarrow$ begin with many quasiparticles.  Owing to the presence of the mobility gap, $\tilde\downarrow$ traces presumably correspond to removing quasiparticles from the heavily quasiparticle-doped $\nu=1/3$ vacuum, while $\tilde \uparrow$ traces correspond to removing quasiholes from the heavily quasihole-doped $\nu=1/3$. The difference between these regimes is highlighted by the asymmetry in the experimentally determined phase slip magnitude. 
As shown in the insets of Figs.\ref{fig:fig6}b and c, while the phase slip magnitudes for $\tilde \uparrow$ traces are clustered around the expected, quantized value $\Delta\theta = 2\pi/3$, for the $\tilde \downarrow$ data the value is instead near $\Delta\theta/2\pi = -0.43$ (Fig. \ref{fig:fig6}c).
Notably, the standard deviation of $\Delta\theta$ in Fig. \ref{fig:fig6}c, where the phase slips are not consistent with the expected quantized values, is no higher than in Figs. \ref{fig:fig2}a or \ref{fig:fig6}b, where $\Delta\theta\approx 2\pi/3$. 

The discrepancy between the observed phase slip magnitude for $\tilde \downarrow$ and the expected quantized value calls into question the interpretation of these slips in terms of the anyon statistical phase $\theta_a$. Deviation from a quantized value 
(e.g., $\theta_a = 2\pi/3$ for $e/3$ anyons, or $\theta_a=2 \pi$ for whole electrons) is most naturally accounted for by invoking bulk-edge coupling.  We consider two scenarios.
In the first, phase slips correspond to the changing the interferometer charge by  $e/3$. In this case, the quantized phase slips observed for rising $\bar{V_C}$ correspond to addition of single anyons to states with negligible bulk-edge coupling, so that the entire phase slip is generated by $\theta_a$.  Decreasing $\bar{V_C}$ then must change the anyon number closer to the edge, where bulk-edge coupling effects are strong.  Alternatively, the phase slips could be \textit{entirely} generated by bulk-edge coupling as would be the case if each event corresponds to the addition of a whole electron, with no observable contribution from $\theta_a$. In either scenario, the asymmetry in the measured values of $\Delta\theta$ indicate that the added charge induced for rising and falling $\bar{V_C}$ must occupy different orbital states with contrasting bulk-edge coupling.

Before analyzing these scenarios quantitatively, we note that the tight clustering of the observed phase slips magnitudes already imposes constraints on the underlying microscopic picture. In particular this observation is inconsistent with a picture where quasiparticles are added to pinning sites randomly distributed throughout the bulk, which would result in a large variation in the bulk-edge coupling and consequently the phase slip values.  
Instead, the tight clustering of phase slip values is more consistent with a picture where quasiparticles are added to a compressible ``puddle'' of charge, so that successive addition of quantized charges produces phase slips of nearly the same magnitude. Notably, our previous study of this regime at $\nu=-1$ in a device of similar construction found phase slips with magnitudes $\Delta \theta \leq 0.1$, considerably smaller than reported here \cite{samuelson_hard_2025}.  

The phenomenological theory of bulk-edge coupling\cite{halperin_theory_2011} gives the phase slip magnitude $\Delta\theta$ in terms of the variables $K_I$ and $K_{IL}$, which parameterize the energetic cost to add charge to the interferometer edge and the cross coupling between the edge and the bulk, as 
\begin{equation}
    \Delta\theta = 2\pi e^* (1-K_{IL}/K_I).
\end{equation}
where $e^*=q/e$.  $K_I = h v_{edge}/L \nu$ can be determined by measuring the change in common mode voltage applied to both the source and drain electrodes simultaneously required to induce a $\Delta \theta = 2\pi$ phase shift; we find $K_I=234\mu eV$ (see Supplement).  
$K_{IL}$ can be estimated separately using a simple microscopic model of a compressible puddle, separated by a distance $l$ from the edge state (see Supplement). 
$K_{IL}$ is maximal when the bulk puddle reaches all the way to the edge, and decreases with increasing separation between edge and bulk. Within this model, the upper bound on $K_{IL}$ is half the charging energy of the interferometer bulk, 
\begin{equation}
K_{IL}\leq \frac{1}{2}\frac{e^2}{C_{bulk}^{g}}
\label{eq:bulkedge}
\end{equation}
where $C_{bulk}^{g}$ is the total geometric capacitance of the interferometer bulk. 
We estimate $C_{bulk}^{g} \approx 2 \epsilon A/d$, where $d$ is the average distance to the gates and $\epsilon\approx 3.25 \epsilon_0$ is the c-axis dielectric constant of hBN, from which we determine an upper bound, $K_{IL}\leq 80\mu eV$.  

Given our estimated $K_I$, then, $K_{IL}$ may be determined from the measured $\Delta\theta$ to check if it is consistent with this bound. For the case of adding  an integer electron charge, $q=e$, the measured $\Delta \theta/2\pi =0.334$ and $-.434$ give $K_{IL}=155\mu eV$ and 132$\mu eV$ respectively, both of which violate the bound.  Our data are thus incompatible with a model of a compressible puddle to which integer electron charges are added.  Notably, this contrasts with a similar analysis of data from integer filling, where this model is found to be in agreement with observations\cite{samuelson_hard_2025}.

In a picture where the compressible puddle is composed of $e/3$ quasiparticles, 
the quantized slips of Fig.~\ref{fig:fig6}b for the $\tilde\uparrow$ data imply a negligible bulk-edge coupling. 
This is consistent with prior results at $\nu=-1$\cite{samuelson_hard_2025}, which found $(1-K_{IL}/K_I) \approx0.9$. Comparable bulk-edge coupling would lead to an expected $\Delta\theta\approx 0.3$ for anyons at $\nu=1/3$.  
However, a crucial feature of the bulk-edge coupling is that in general it only \textit{reduce} the observed phase slip magnitude associated with the entry of a single quasiparticle.  Thus the phase slips with $\Delta \theta = -0.43$ in Fig.~\ref{fig:fig6}c for the $\tilde \downarrow$ sweeps cannot be accounted for in a model where each slip corresponds to the addition of only one $e/3$ quasiparticle, even near the sample boundary.  

Thus, while the non-quantized value implies that bulk–edge coupling must play a role in determining the phase slip magnitudes, it is also evident that neither a model based on whole-electron jumps nor single-$e/3$ quasiparticle jumps can account for the observed slips. This motivates consideration of alternative processes, such as those involving multiple quasiparticles. In particular, correlated dynamics—whereby quasiparticles may enter or exit in pairs—could give rise to phase slips with magnitudes distinct from those expected for individual particles. In the following section, we identify evidence for such processes by analyzing the fine structure of individual interference patterns---i.e., features beyond the primary phase---encoded in modulations of the fringe amplitudes.

\section{Time-Domain Reconstruction of a 2$e$/3 Jump}

\begin{figure}
    \includegraphics[width = 88mm]{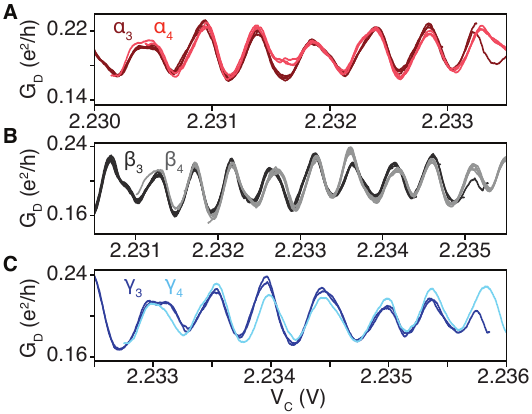}
    \caption{
    \textbf{Comparison of charge fingerprints for quantized phase slips.}
    Traces are shown in the gate voltage range where they overlap, corresponding to the classes in Fig. \ref{fig:fig5}A labeled \textbf{(A)} $\alpha_3$ and $\alpha_4$, \textbf{(B)} $\beta_3$ and $\beta_4$, \textbf{(C)} $\gamma_3$ and $\gamma_4$.
\label{fig:fig7}}
\end{figure}

Within an individual panel of Fig. \ref{fig:fig5}a, all curves are measured under identical conditions of magnetic field and applied gate voltages. As might be expected, traces taken at different times prior to a charging event show excellent reproducibility. A less obvious question is whether the addition of charge $q = e$ to the interferometer, particularly when it occurs deep in the bulk, modifies the $G_D-V_C$ relation even though no effect on the interferometer phase is expected. In figure \ref{fig:fig7}a-c, we collect traces from the data of Fig. \ref{fig:fig5}a which are separated by three quantized phase slips, and indeed, see no discernible change in the oscillation \textit{phase}.
Notably, however, despite accumulating $\Delta \theta = 2\pi$, traces which represent charge configurations differing by a whole electron---e.g., $\alpha_3$ and $\alpha_4$---remain distinguishable through deviations in the precise \textit{amplitude} of individual interference fringes.
This effect is robust across nearly all groups of traces separated by a $2\pi$ phase shift (see Supplement for the complete set).  We speculate that this `charge fingerprint' encoded in the fringe intensities arises from the exponential sensitivity of transmission through our quantum point contacts to their electrostatic environment---including the electrical potential generated by the addition of even a gate-screened electrical charge in the bulk. 

\begin{figure}[hb!]
    \includegraphics[width = 88mm]{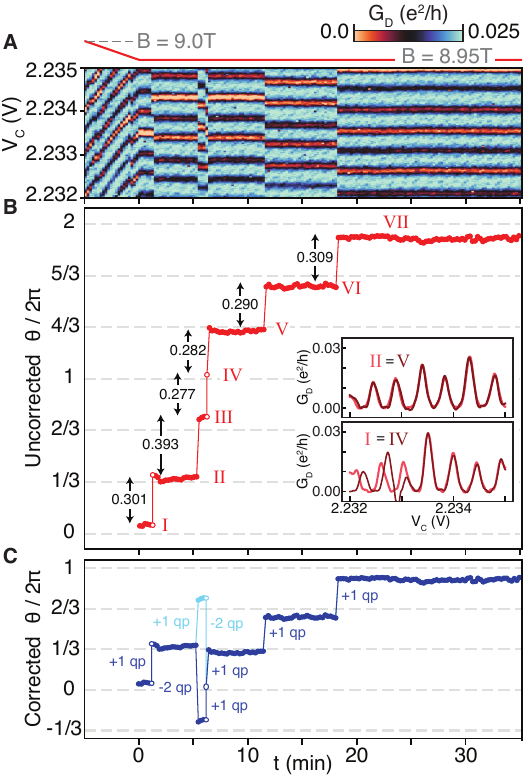}
    \caption{
\textbf{Time domain reconstruction of the quasiparticle occupation.}
\textbf{(A)} Repeated measurement of $G_D$ over a fixed window of $V_C$. At the beginning of the measurement, the magnetic field is swept from 9.0T to 8.95T, where it is subsequently held constant for 35 min.
\textbf{(B)} The phase, $\theta$, extracted from the Fourier transform of the data in panel A. Each jump is assumed to be in the interval $\Delta \theta \in (-\pi, \pi)$. For traces in which a phase jump is observed in the middle of the line trace, two points are plotted as open circles, corresponding to the two values of phase extracted before and after the jump. Inset: $G_D$ traces from regions II and V, vs. compared to line traces taken from regions I and IV. \textbf{(C)} Two versions of the corrected phase, $\theta$, identical to panel B except with a shift of $-2\pi$ applied to the jump from II to III (dark blue), or the jump from III to IV (light blue). Either modification would match the 'charge fingerprints' of regions I and IV, and regions II and V.
\label{fig:fig8}}
\end{figure} 

Regardless of microscopic origin, however, the pattern of oscillation amplitudes provides information about the charge configuration not contained in the oscillation phase alone.  As a result, it allows us to extract information about the absolute phase (interpreted as the total electron charge in the interferometer\cite{feldman_robustness_2022}) rather than just the phase modulo $2\pi$.  
This in turn allows us to resolve quasiparticle \textit{entry} and \textit{exit} under static experimental conditions in the time domain.  

Fig. \ref{fig:fig8}a shows the result of ramping the magnetic field to a set point and then, starting from $t=0$, leaving it constant for 35 minutes while continuing to repeatedly ramp $V_C$ over a small ($\sim \SI{1.5}{mV}$) range. Several phase slips are visible over the course of the experiment, separating regions of stable phase which we label I-VII in Fig. \ref{fig:fig8}b.
If the phase jumps were assumed to fall into the interval $(-\pi, \pi)$, their nearly consistent magnitudes could lead to the conclusion that every jump corresponds to the addition of one quasiparticle, as is energetically favored by the reduction in total flux at the beginning of the measurement.

However, a close examination of the measured conductance (Fig. \ref{fig:fig8}b insets) shows the fringe intensity patterns in regions II and V to be indistinguishable; regions I and IV are also identical after noting that the phase slip occurs midway through the single trace that comprises region IV. This leads us to identify these regions with a return to the same charge configuration as well as the same absolute phase. In fact, equating the regions which have an indistinguishable `charge fingerprint' necessitates that at least one of the first 3 phase slips must correspond to a \textit{removal} of two quasiparticles, rather than the na\"ive interpretation of every slip as the addition of a single quasiparticle. Fig. \ref{fig:fig8}C shows two possible corrected traces of $\theta(t)$, with the light and dark blue traces differing from the uncorrected $\theta$ by a shift of $-2\pi$ at the second or third phase slip, respectively. 

The time dependence of the quasiparticle number implied by either of these possibilities reconciles our earlier observation of the identical `charge fingerprints' in \textit{both} regions I and IV, and regions II and V.
The information required to discriminate between the two possibilities is provided by the magnitude of the phase slip from region II to region III: here, $\Delta \theta$ is significantly \textit{larger} than $\theta_a$, whereas a finite amount of bulk-edge coupling should only serve to decrease the magnitude of the phase jump. (All of the other phase slips show a value slightly less than $\theta_a$). This suggests that the slip  separating regions II and III should be associated with a decrease in the number of quasiparticles by two, as the change in phase is then $\Delta \theta = -4\pi / 3 + \delta \theta_{BE}$. In other words, we interpret $\Delta \theta / 2 \pi = 0.393 = -\frac{2}{3}(1 - K_{IL}/K_{I}) +1$, which gives a  value of $K_{IL}/K_{I} \approx 0.1$ consistent with small 
bulk-edge coupling.

In addition to the potential for such processes to account for the non-quantized phase slips from Fig.~\ref{fig:fig5}c, we note that double-quasiparticle tunneling is not without precedent. Experimentally, an `Andreev'-like scattering process involving a charge-$2e/3$ excitation at the $\nu = 1/3$ edge has been observed in a $\nu = 1$ to $\nu = 1/3$ heterojunction \cite{cohen_universal_2023}, and shot noise measurements in the fractional quantum Hall regime have sometimes found evidence of cooperative tunneling of multiple quasiparticles \cite{comforti_bunching_2002, bid_shot_2009, ghosh_coherent_2025}. Capacitive signatures have also revealed correlated double-electron additions to quantum dots in the integer quantum Hall regime \cite{demir_correlated_2021}. Furthermore, recent theoretical work has suggested that the lowest energy multi-quasiparticle configuration in some fractional quantum Hall systems may involve formation of ``anyonic molecules'' or clusters composed of two or more quasiparticles \cite{gattu_molecular_2025, xu_dynamics_2025}.

\section*{Discussion}
The exceptionally slow quasiparticle dynamics of our interferometer open new routes to quantitatively probe the physics of fractionalized phases at the single anyon level. 
Measuring few-anyon dynamical processes via the response of the interferometric phase to both $\theta_a$ and Coulomb effects may give new insight into states where inter-quasiparticle correlations are important, such as in the formation dynamics of Wigner crystal states, anyonic molecules, and the hierarchical fractional quantum Hall states.

Recent advances in spatial imaging techniques, allowing high resolution charge sensing in dual-gated devices\cite{li_imaging_2021}, may be used to help resolve these questions. Applied to interferometer devices of a similar construction to ours, this may allow direct correlation between the real-space distribution of localized anyons and the interferometric phase. Graphene heterostructures also host even denominator fractional quantum Hall states thought to support non-Abelian anyons\cite{ki_observation_2014,zibrov_tunable_2017,li_even_2017}. In bilayer graphene, the energy gaps of these states are comparable to that of the 1/3 state studied here \cite{assouline_energy_2024,hu_high-resolution_2025}, suggesting similarly slow dynamics for charge e/4 quasiparticles. A key open question is the timescale for motion of the charge-neutral excitations that encode fermion parity, which future experiments of this kind may help illuminate.

\textit{Note Added:} During the preparation of this manuscript, we became aware of related work using a similar graphene device \cite{werkmeister_anyon_2025}.

\section*{Data Availability}
The data that support the findings of this study are available from the corresponding author upon reasonable request.

\section*{Acknowledgements} 
The authors acknowledge helpful discussions with T. Wang, S. Kivelson and S. Das Sarma. Work at UCSB was primarily supported by the Office of Naval Research under award N00014-23-1-2066.  Development of fabrication based on anodic oxidation lithography were supported by the Air Force Office of Scientific Research under award FA9560-20-1-0208. 
AFY acknowledges additional support by the Gordon and Betty Moore Foundation EPIQS program under award GBMF9471. 
LC and NS received additional support from the Army Research Office under award W911NF20-1-0082. 
MZ was supported by the U.S. Department of Energy, Office of Science, Office of Basic Energy Sciences, Materials Sciences and Engineering Division under Contract No. DE-AC02-05-CH11231 (Theory of Materials program KC2301).
K.W. and T.T. acknowledge support from JSPS KAKENHI (Grant Numbers 19H05790, 20H00354 and 21H05233).

\bibliographystyle{apsrev4-1}
\bibliography{references}
\clearpage
\pagebreak
\setcounter{page}{1}

\title{Supplementary Information}
\maketitle
\onecolumngrid
\renewcommand\thefigure{S\arabic{figure}}
\setcounter{figure}{0}
\setcounter{equation}{0}

\section{Materials and Methods}

\subsection{S1. Sample Fabrication}
\begin{figure}[hb!]
    \centering
    \includegraphics[width = 120mm]{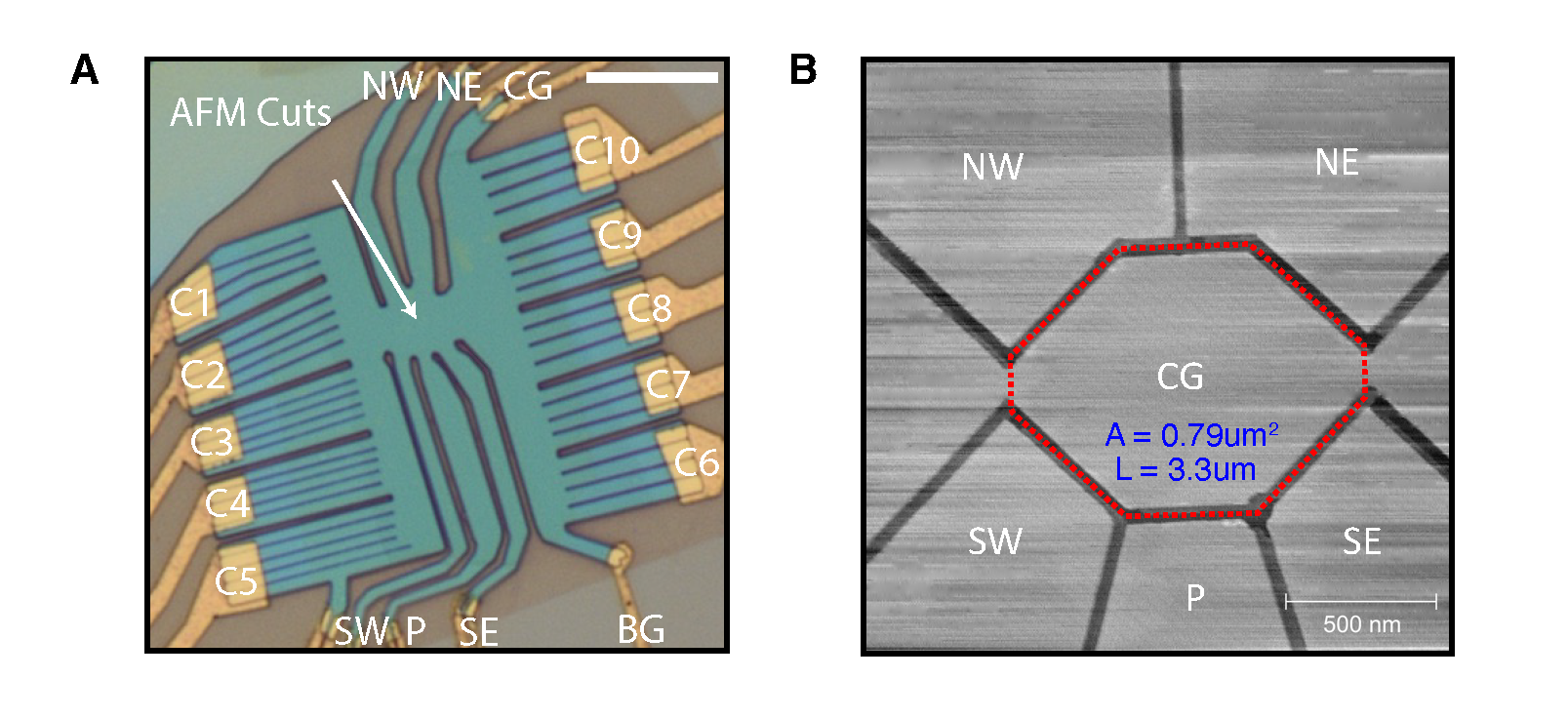}    \caption{\textbf{Device Image and AFM Cut Gate} \textbf(a) Optical micrograph of the device.  Electrical edge contacts to the monolayer are labeled as C1-10.  Electrical edge contacts to the gate layers, NW, SW, NE, SE, P, CG, and BG are marked respectively. Scale bar inset is $\SI{10}{\mu m}$.  \textbf{(b)} AFM topograph on a transfer slide of the top gate layer after pick-up with an hBN flake.  The gate regions shown in panel a, as well as in Fig.~\ref{fig:fig1}a, are marked.  The measured lithographic area, defined by the perimeter drawn in the red dashed line, is measured to be $\SI{0.79 \pm  0.05}{\mu m^2}$, and the perimeter is measured to be $\SI{3.3 \pm 0.1}{\mu m}$.  The uncertainty comes from the width of the etched region in the graphite top gate (it is unknown exactly where the edge state is located for a particular set of gate voltages). } 
\label{fig:device_image}
\end{figure}

The sample (optical image shown in Fig. \ref{fig:device_image}A) was fabricated using the standard van der Waals dry-transfer process combined with electrode-free AFM anodic oxidation lithography of the graphite top gates before stacking, as described in reference \cite{cohen_nanoscale_2023}. The structures patterned into the top graphite gate are shown in the AFM topograph of Fig. \ref{fig:device_image}B. The scan was acquired in the middle of the stacking process, imaging the exposed graphite top gate after picking it up. After stacking, the device is transferred onto a conductively-doped Si substrate with \SI{285}{nm} of thermally-grown SiO$_2$ on the surface.

A mask is defined via E-Beam Lithography (EBL) of a polymethyl methacrylate A4 495K/A2 950K bilayer resist, and \SI{40}{nm} of aluminum is deposited onto the device to be used as a hard etch mask. The whole stack is then etched through using a CHF$_3$/O$_2$ plasma reactive ion etch for several minutes. The aluminum mask is removed by submersion in a $<$3\% TMAH solution (AZ300MIF) for 20 minutes. Then, a contact electrode pattern is defined using a PMMA 950K A8 mask and another EBL exposure, a 30 second CHF$_3$/O$_2$ RIE etch is performed to clean the expose device edges, and edge contacts are deposited by E-Beam evaporation of 5/15/150nm Cr/Pd/Au layers.

\subsection{Voltages for Main Text Figures}
\begin{figure}[H]
    \begin{center}
    \begin{tabular}{||c c c c c c||} 
     \hline
     \textbf{Main Text Figure} & $V_{CG}$ & $V_{BG} $ & $V_{NW/SW}$ & $V_{NE/SE}$ & $V_P$ \\ [0.5ex] 
     \hline\hline
     Fig.~\ref{fig:fig1}c (blue trace) & 2.23V & -2V & 2.23V & -4.1V & 2.23V \\ 
     \hline
     Fig.~\ref{fig:fig1}c (red trace) & 2.23V & -2V & 0.5V & 2.23V & 2.23V \\
     \hline
     Fig.~\ref{fig:fig2}a & 2.23V & -2V & 0.5V & -4.1V & N/A \\
     \hline
     Fig.~\ref{fig:fig3}a-b & N/A & -2V & 1.0V & -3.2V & 2V \\
     \hline
     Fig.~\ref{fig:fig5},\ref{fig:fig6},\ref{fig:fig7} & N/A & -2V & 1.0V & -3.2V & 2V\\
     \hline
     Fig.~\ref{fig:fig8} & N/A & -2V & 0.3V & -3.94V & 2V\\ [1ex]
     \hline
    \end{tabular}
    \end{center}
    \caption{\textbf{Fixed voltage set points for main text figures}}
\end{figure}

\subsection{S2. Measurement parameters and interferometer characterization}

Experiments were performed in a dry dilution refrigerator at a base temperature of \SI{18}{mK}. Electronic filters are installed on all transport lines to lower the electron temperature. Transport was measured at \SI{10.3875}{Hz} using SR860 lock-in amplifiers. A Basel Precision Instruments SP983c high stability I to V converter (IF3602) is used to measure the current signals, while the voltages are measured directly with the SR860 using no additional pre-amplification.

For the diagonal conductance measurements an ac excitation is applied to contacts C6, C7, C8, and C9 and the resulting current $I_{out}$ is measured on contact C1. The diagonal voltage drop $V_{+} - V_{-}$ is measured between contacts C10 and C2. A constant \SI{10}{V} is applied to the conductively-doped Si substrate throughout all measurements to maintain electron-doping in the single- and un-gated regions of each contact. The excitation voltage is $\SI{10}{\mu V}$ for the data in Fig. \ref{fig:fig2}a and Fig. \ref{fig:fig8}, and $\SI{20}{\mu V}$ for Fig. \ref{fig:fig3}A-B, and Fig. \ref{fig:fig5},\ref{fig:fig6},\ref{fig:fig7}.

As described in the main text, we perform several standard characterizations of our Fabry-Perot interferometer to confirm normal operation, shown in the figures that follow. 

\begin{figure}[H]
    \centering
    \includegraphics[width=0.7\textwidth]{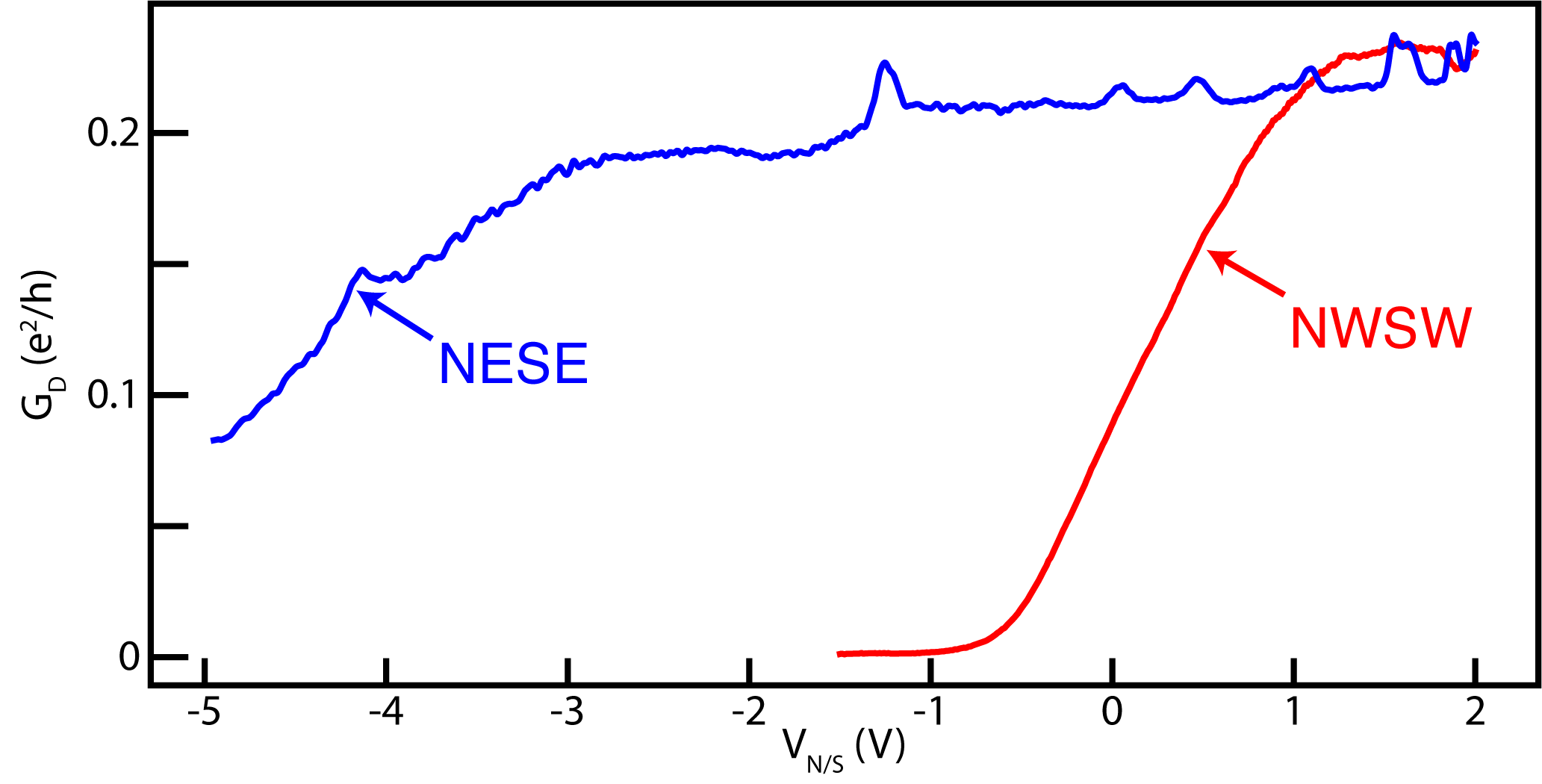}
    \caption{\textbf{Individual QPC pinch-off curves in $\nu = 1/3$.}  Blue curve is $G_D$ vs. $V_{NE/SE}$ while $V_{NW/SW}$ is set such that $\nu_{NW/SW} = 1/3$.  Red curve is $G_D$ vs. $V_{NW/SW}$ with $V_{NE/SE}$ set such that $\nu_{NE/SE} = 1/3$.  Operating points for both QPCs relevant to the interference data shown in Fig.~\ref{fig:fig2}a are identified by the inset arrows.  A large asymmetry in the NESE vs. NWSW QPCs is observed; this may be understood as arising from an intrinsic asymmetry in the lithography defining the constrictions that set the width of the NESE and NWSW QPCs.  Since a large voltage is required to pinch-off the transmission in this geometry, the NS regions develop additional co-propagating edge modes that screen the applied gate potential before pinch-off is achieved.  This screening reduces the capacitive coupling between the QPC saddle point and the N/S gates, leading to the large asymmetry in voltage observed between the E and W QPCs.}
\label{fig:qpc_pinch_off}
\end{figure}

\begin{figure}[H]
    \centering
    \includegraphics[width = 0.95\textwidth]{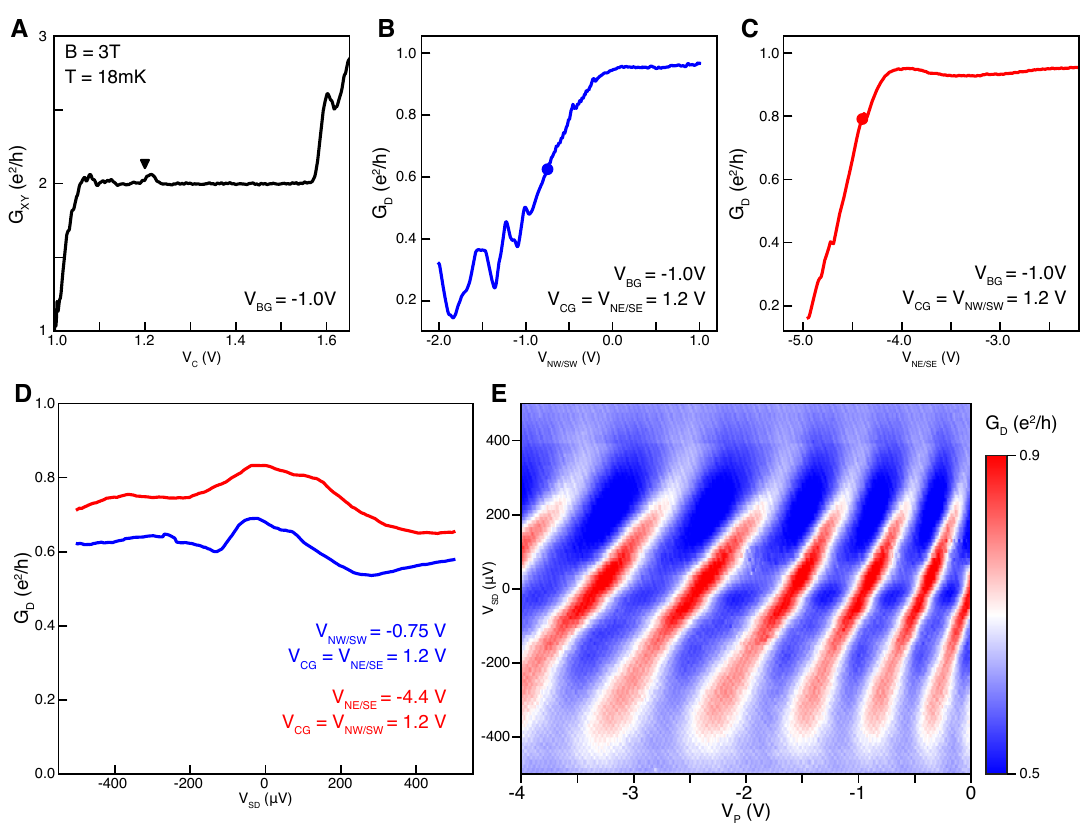}
    \caption{\textbf{Bias Dependence of the QPC transmission in the outer edge of the $\nu = 2$ state.} \textbf{(A)} Hall conductance as a function of $V_C$ at 3T. The interferometer is characterized in the $\nu = 2$ state with the center gate fixed at the marked point. \textbf{(B, C)} Diagonal conductance across the device with each QPC pinched off individually. \textbf{(D)} Dependence of the transmission through each QPC at the marked points in B,C on the applied source-drain DC bias $V_{SD}$. No zero-bias suppression in the conductance is observed, in stark contrast to the phenomenology observed in the FQH regime. \textbf{(E)} Dependence of the interference on the source drain bias and plunger gate, with both QPCs set to the marked points in B,C simultaneously.}
\label{fig:supp_bias_dep_int_qpc}
\end{figure}

\begin{figure}[H]
    \centering
    \includegraphics[width = 0.8\textwidth]{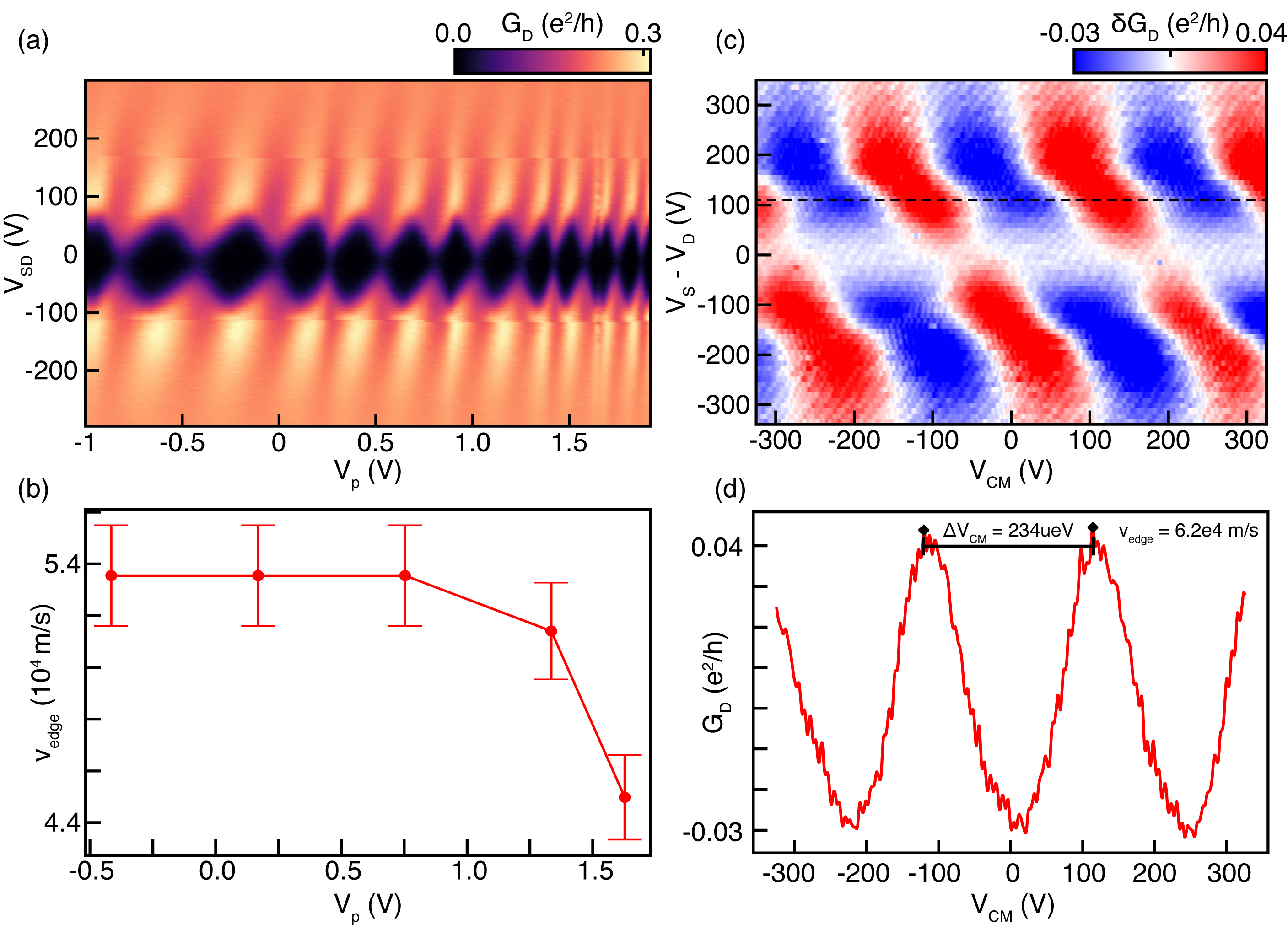}
    \caption{\textbf{Bias and common mode dependence of interference in $\nu$ = 1/3} \textbf{(a)} $G_D$ versus $V_P$ and $V_{SD}$ near $\nu = 1/3$, B = 9T, $V_{CG} = 2.27V$, $V_{NW/SW} = -0.23V$, $V_{NESE} = -4.25V$.  The interference shows a strong suppression of $G_D$ near zero DC bias, characteristic of chiral Luttinger liquid behavior near the QPCs.  Additionally, the interference exhibits a phase-shift resulting from an applied $V_{SD}$.  Notably, the periodicity in $V_{SD}$ is reduced for values of $V_P < 0.5V$.  \textbf{(b)} Extracted Edge velocity from the periodicity in $V_{SD}$ in panel a versus $V_P$, here we use the formula $h v_{edge} / \nu L = e\Delta V_{SD}$.   As $V_P$ becomes enhanced the edge velocity exhibits a 19\% reduction coming from the decreasing sharpness of the confinement potential near the plunger gate.  \textbf{(c)} $\delta G_D$ as a function of both the common mode voltage on the monolayer, $V_{CM}$, and the difference between the voltages on the source and drain terminals relative to fridge ground, $V_S - V_D$; this difference is taken by varying the voltage on $V_S$ while keeping $V_D$ fixed.  This data is taken in the same operating regime as Fig.~\ref{fig:fig1}d.  The periodicity in $V_{CM}$ directly tells us e/$C_I$, and consequently the edge velocity as quoted in the main text.  \textbf{(d)}  Line cut along the dashed black line in panel c.  We can extract the peak-to-peak distance between fringes versus $V_{CM}$, which yields a periodicity $\Delta V_{CM} = \SI{234}{\mu V}$.  Using the relation $h v_{edge} / \nu L = e\Delta V_{CM} = e^2 / C_I$, we find that $v_{edge} = \SI{6.2e4}{m/s}$.  Despite the fact the panels a-b and panels c-d are taken at slightly different gate voltages we observe good agreement with the predicted edge velocity for the $\nu = 1/3$ edge mode using two different methods.}
\label{fig:supp_bias}
\end{figure}

\clearpage
\newpage

\subsection{S3. Bulk-edge and exterior-edge coupling at $\nu = \frac{1}{3}$}

\emph{The edge ``capacitance.''}
In the main text we reported the common-mode periodicity $\Delta V_{CM} = \SI{234}{\mu V}$, which we relate to the capacitance of the edge as $e = \Delta V_{CM} C_I$.  This measurement is taken in the regime where the bulk is completely incompressible (no phase slips). 
But we should note that in the notation of Halperin et al. \cite{halperin_theory_2011}, this measurement is  more properly understood as a characterization of $K_I = \SI{234}{\mu e V}$. When the bulk is entirely incompressible, the relation between the two is trivial: $K_I = e^2 / C_I$. However, if extended to the regime when the bulk is compressible (for example at a phase slip), and with finite-bulk-edge coupling $K_{IL}$, the parameters are related as a matrix inverse $K = e^2 C^{-1}$, so that $C_I = \frac{e^2}{K_I  - K^2_{IL}/ K_L}$. 
We do not expect $K_I$ to depend too much on the bulk filling, while $C_I$ will depend on the filling via quantum capacitance effects in $K_L$. 
Thus when we refer to  $C_I$ as ``the capacitance of the edge,'' strictly speaking this should be understood as $e^2 / K_I$.
In any case the capacitance matrix is not so relevant to our experiments because the bulk and edge are out of equilibirium.

\emph{Bulk-edge coupling of the anyonic phase.}
We  elaborate on the derivation of Eq.\eqref{eq:bulkedge} of the main text, which gives the Coulomb coupling between an anyonic impurity and the edge. Assuming the radius of curvature of the edge is large compared with the impurity distance, we can treat the edge as a line with capacitance per unit length $C_I / L$. The induced charge is then $\delta Q_I =  \frac{C_I}{L} \frac{e}{3} \int  G(x, R) dx$ where $G$ is the double-gated Green's function with a charge $e/3$ at $(x, y) = (0, R)$.
Making use of the effective translation variance along $x$, we can then use a conformal transformation to  find that a 1D line charge $\rho$ produces a potential $\phi(y) = \frac{\rho}{2 \pi \sqrt{\epsilon_{xy} \epsilon_z}}\log(\tanh(\pi y \sqrt{\epsilon_z / \epsilon_{xy}} / 4 d)) \approx \frac{1}{\pi \sqrt{\epsilon_{xy} \epsilon_z}} e^{-\pi y \sqrt{\epsilon_z / \epsilon_{xy}} / 2 d}$. This gives the estimate $\delta Q_I =  \frac{e}{3}  \frac{C_I}{\pi \sqrt{\epsilon_{xy} \epsilon_z} L } e^{-\pi y \sqrt{\epsilon_z / \epsilon_{xy}} / 2 d}$.
This also provides an estimate of the \emph{average} bulk-edge coupling, $K_{IL} = \frac{e^2}{A} \int_{y>0} \int dx G(x, y) dx  dy = \frac{e^2}{2 A} \frac{d}{2 \epsilon_z} = \frac{e^2}{2 A c_g}$, where $c_g$ is the geometric bulk capacitance per unit area. 

This discussion also points out the need to generalize the analysis of Ref.\cite{halperin_theory_2011} in order to describe  bulk-edge coupling in our experiments. In that work, the bulk was assumed to be in electro-chemical equilibrium and that it could be treated as a single lumped element with charging energy $K_{L}$. In our experiment, however, we may be resolving charging of single impurities. To explain bulk-edge coupling in this regime  requires a more fine-grained  model keeping the charge state of each impurity $a$ and an impurity-edge coupling $K_{L, a}$, which is precisely the quantity analyzed above. 
\vspace{4mm}

\emph{Bulk-edge coupling of the AB phase.}
In Fig. \ref{fig:fig2} of the main text we found that in the regime of Aharonov-Bohm oscillations in which $N_{qp}$ is kept fixed, the field period $\Delta B$ varies from $\SI{15}{mT}$ to $\SI{18}{mT}$ as the plunger gate voltage $V_P$ is varied from  $\SI{-0.75}{V}$ to $\SI{-2.25}{V}$.
Assuming pure Aharonov-Bohm evolution,  $\frac{e}{3} A \Delta B  = h$, this gives a range of areas $A \sim 0.69 - \SI{0.83}{\mu m^2}$ in good agreement with the nominal geometric area of the device.  Here we discuss two qualifications to this analysis, and their possible resolution: 
\begin{itemize}
    \item The measured \textit{variation} in  $\Delta B$ is larger than can be explained by the $V_P$-dependence of $A$. 
    \item  In the standard theory of bulk-edge coupling,\cite{halperin_theory_2011} the field period should  receive a correction $\frac{e}{3} \bar{A} (1 - K_{IL} / K_I) \Delta B  = h$, and from theoretical estimates of the $K_{IL}$ it is  surprising the field period is in such good agreement with the nominal area. 
\end{itemize}

We propose that the resolution to both arises from a novel feature of our graphene Fabry-Perot interferometer geometry, which has not been considered in the literature focused on GaAs devices.  Specifically, the filling outside the interferometer is of opposite sign to the interior ($\nu_{\textrm{ext}} \ll 0$ for positive filling of the interferometer), 
giving rise to coupling between the edge state and this exterior region we term ``exterior-edge'' coupling.  Exterior-edge coupling counteracts the effect of the typical bulk-edge coupling between the edge state and the interior of the interferometer. 

The variation in $\Delta B$ would at first seem to be intuitive, as $\theta = \frac{e}{3} A_I(V_P, B) B / \hbar$ and $V_P$ increases the area $A_I$.
However, the observation of only 4 interference periods over this range of $V_P$ implies $A_I$ has changed by an area of  $ 4 \cdot 2 \pi \ell_B^2 / \nu \sim \SI{e-3}{\mu m^2}$, which is negligible compared to the observed $20\%$ variation in $\Delta B$.
We conjecture the $V_P$ dependence instead occurs through the $B$-dependence of $A_I$,  $\partial_B \theta = \frac{e}{3} A_I ( 1 + B A_I^{-1} \partial_B A_I)  / \hbar$.
This dependence is one manifestation of bulk-edge Coulomb coupling: as $B$ is increased at fixed $N_{qp}$ (and hence $\nu$),  the electron density $n = \nu e B / \hbar$ in the bulk increases, which in turn tends to shrink the edge due to Coulomb coupling, reducing $A_I$.
The standard theory \cite{halperin_theory_2011, von_keyserlingk_enhanced_2015} of  bulk-edge coupling  predicts
\begin{align}
\label{eq:bulk_edge_coupling}
\theta = (1 - \frac{K_{IL}}{K_I} \frac{\nu_{\textrm{in}}}{\Delta \nu} ) e^\ast \bar{A} B / \hbar      
\end{align}
where in the present case $\nu_{in} = \Delta \nu = 1/3, e^\ast = e/3$. 
Here $K_I \sim e^2 / C_I \approx \SI{234}{\mu eV}$ describes the compressibility of the edge and $K_{IL}$ is a phenomenological bulk-edge Coulomb coupling. 
The change in $\Delta B$ could then arise from a $V_P$ dependence of $K_I$ or $K_{IL}$. 
Based on the electrostatics of a double-gated device, we expect bulk-edge coupling of $K_{IL} = e^2 / 2 c_g A \approx \SI{86}{\mu eV}$, where $c_g = 2 \epsilon_z / d$ is the capacitance to the gates. 
Furthermore, we have measured $K_I \sim \SI{234}{\mu eV}$ and find it changes by only about 15\% as $V_P$ is varied over an even larger range of $V_P$ than displayed in Fig 1. 
The prediction of the standard theory then raises a puzzle, as the resulting correction by $(1 - 86 / 234) \approx 0.63$ would imply that the actual area is in fact $\sim 60\%$ larger than previously inferred, $\bar{A} \sim 1.1 - \SI{1.3}{\mu m^2}$.
This is at odds with the nominal size of the interferometer $0.74 - \SI{0.83}{\mu m^2}$.

We conjecture this discrepancy arises from an effect not previously considered in the context of GaAs FPI. 
In the standard theory of bulk-edge coupling, it is implicitly assumed that $\nu \to 0$ within a short distance of the edge.
In our graphene FPI, however, $\nu \sim -5$ in the exterior of the interferometer.
As a result we expect additional Coulomb coupling between the charge in the \textit{exterior} of the FPI and the edge. When the exterior is operated in a gap, the exterior charge  density will introduce a new  term to the Coulomb correction: $(1 - \frac{K_{IL}}{K_I} \frac{\nu_{\textrm{in}}}{\Delta \nu} - \frac{K_{I, \textrm{ext}}}{K_I} \frac{\nu_\textrm{ext}}{\Delta \nu} )$.
In  double-gated devices, we again estimate  $K_{I, \textrm{ext}} \sim K_{IL} \sim  e^2 / 2 c_g A$ (while it may seem strange for $K_{I, \textrm{ext}}$ to depend on $A$, this is an artifact of the factor of $A$ which has been factored out in Eq.\eqref{eq:bulk_edge_coupling}: for both the bulk and exterior coupling, the important scale is the  gate distance relative to the perimeter,  $d / L$). 
For $\nu_\textrm{ext} < 0$ the exterior charge will thus counteract the bulk-edge coupling. 
In double-gated devices, fields are screened over a scale $d$, so the relevant $\nu_\textrm{ext}$ is one averaged over a distance of $\sim d$ from the interfering edge. Over this range we expect the density to fall from $\nu = \frac{1}{3} \to -5$ through a series of steps whose spatial structure depends on the details of the electrostatics and interaction energies.
It is thus  difficult to quantitatively predict $\nu_\textrm{ext}$. 
Nevertheless, this hypothesis can be be tested through a future study of the interferometer field period as a function of $\nu_{\textrm{ext}}$, though this requires a modified device design. 

\clearpage
\newpage

\subsection{S4. Temperature dependence and thermal dephasing of the interference phase in the FQH regime}
\begin{figure}[H]
    \centering
    \includegraphics[width = 184mm]{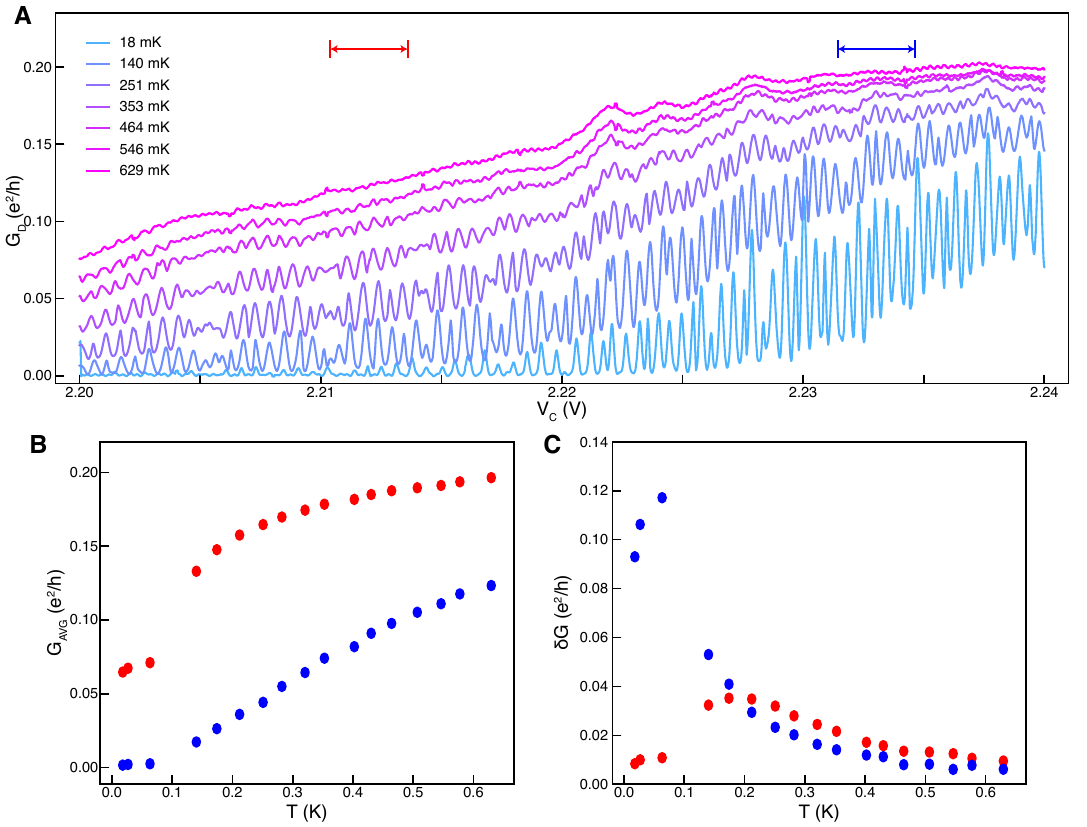}
    \caption{\textbf{Temperature dependence in FQH} \textbf{(a)} $G_D$ versus $V_C$ in $\nu = 1/3$, with both QPCs at partial pinch off, showing clear oscillations for a range of temperatures measured on our sample probe between $T = \SI{18}{mK}$ and $T = \SI{639}{mK}$.  \textbf{(b)} The average value of the diagonal conductance, $G_{AVG}$, plotted versus the temperature on the probe $T$ for two ranges in $V_C$ marked by the red and blue arrows respectively in panel a.  The average value reflects the tunneling rate of the two QPCs, which for a chiral Luttinger liquid is expected to change significantly with increasing temperature.  \textbf{(c)} Visibility, denoted as $\delta G$ and defined as the average difference between successive maxima and minima in the interference is plotted versus the temperature on the probe, $T$.  Interestingly, the interplay between the chiral Luttinger liquid renormalization of the QPC transmission and the temperature induced decoherence of the interference leads to a maximum in the visibility that depends on the initial set point of $V_C$.   }
\label{fig:FQHTempdep}
\end{figure}

\begin{figure}[b!]
    \centering
    \includegraphics[width = 120mm]{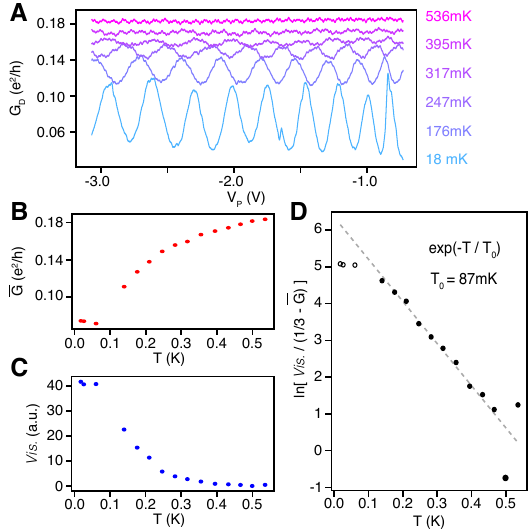}
    \caption{ \textbf{Scaling of the interference visibility with temperature, suggesting a finite amount of thermal dephasing from $N_{qp}$ fluctuations} \textbf{(A)} $G_D$ vs. $V_P$ oscillations in the $\nu = 1/3$ state, measured as a function of temperature at fixed $V_C = \SI{2.230}{V}$. \textbf{(B)} The average value of the conductance $\bar{G}$ as a function of temperature shows  strong temperature dependence,  characteristic of a chiral Luttinger liquid edge. \textbf{(C)} Interference visibility, defined by taking the absolute value of the largest Fourier component of the oscillations at base temperature, and tracking the amplitude of that component as a function of temperature. \textbf{(D)}To cancel the power-law dependence of the backscattering, here we normalize the visibility by
the average backscattering across the interferometer, $(e^2/3 h - \bar{G})$. The normalized visibility  exhibits  exponential suppression $e^{-T / T_0}$ from which we extract $T_0 = \SI{87}{mK}$}
\label{fig:supp_temp_plunger}
\end{figure}

As the temperature increases, the amplitude $\delta G$ of the conductance oscillations is observed to decrease; this is theoretically expected due to dephasing processes. 
In the FQH case, quantitative analysis of this dependence is complicated by the non-linear $T, V$ characteristic of the current: the effective back-scattering amplitude of a \emph{single} quantum-point contact scales as a power-law in $T$ \cite{wen_edge_1991, kane_transport_1992, chamon_two_1997}.  This can lead to a complex behavior of the interference as a function of temperature, where visibility may even exhibit a maximum at finite temperature for a fixed set of operating voltages, as shown in Fig.~\ref{fig:FQHTempdep}C.  At least in the limit of perturbative  backscattering, this RG flow can be cancelled by considering the ratio of the oscillation amplitude and the mean current,
$\delta I_B / \bar{I}_B = \delta G_D / (e^2/3h - \bar{G}_D)$ \cite{halperin_theory_2011}.
The result is shown in Fig. \ref{fig:supp_temp_plunger}A, and over a range of $T$ shows exponential dependence  $e^{-T / T_0}$ with $T_0 = \SI{87}{mK}$.

When accounting for  thermal fluctuations of the conformal field theory on the edge alone, Ref.\cite{chamon_two_1997} predict $k T_0 = \frac{\hbar v}{\pi \nu L}$ (note  their notation for the edge geometry is related to ours by $a = L/2$). 
Relating this to the measured edge capacitance $\frac{h v}{ \nu L} = \frac{e^2}{C_I} \approx \SI{234}{\mu eV}$, we would then predict $k T_0 = \frac{e^2}{2 \pi^2 C_I} = \SI{137}{mK}$, in rather poor agreement with our experiment. 
Here we show the discrepancy with the CFT prediction can be accounted by for by including fluctuations in the bulk quasiparticle number, \cite{halperin_theory_2011} 
an effect also observed in Ref.\cite{nakamura_direct_2020}.

Noting that the interference phase is proportional to the total electron charge $\theta = 2 \pi N$, the  thermal dephasing of $\langle e^{i \theta} \rangle_T$ arises from fluctuations in $N$. 
The result can then be estimated given the capacitance of the edge $C_I$ and bulk $C_L$.\cite{halperin_theory_2011}
Focusing on an operating point where $\langle N_{qp} \rangle_T = 0$, and, for simplicity, ignoring bulk-edge coupling, we obtain 
\begin{align}
\frac{\langle e^{i \theta} \rangle_T}{\langle e^{i \theta} \rangle_0} &=  \left( \frac{1}{Z_I} \int e^{2 \pi i N - \frac{e^2 N^2 }{2 C_I k T} } dN \right) \left( \frac{1}{Z_B} \sum_{N_{qp}} e^{i 2\pi i (N_{qp}/3) -\frac{ (e/3)^2 N^2_{qp}}{2 C_L k T}}\right) \\
&\approx e^{- 2 \pi^2 k T (C_I +C_L) / e^2 } = e^{- T / T_0}
\end{align}
To obtain the exponential approximation we have taken the high-$T$ limit where fluctuations in $N_{qp}$ are large; for $k T < \frac{(e/3)^2}{C_L}$,  $N_{qp}$ fluctuations will be frozen out (or un-observable when the equilibriation time becomes long).
Below we  obtain the estimate $\frac{(e/3)^2}{C_L} \sim \SI{170}{mK}$, and experimentally we indeed find good exponential behaviour above this scale.
As discussed, using the measured value of $\frac{e^2}{C_I} = \SI{234}{\mu eV}$, we find that the edge alone contributes $ k T_0 = e^2 / 2 \pi^2 C_I \sim \SI{137}{mK}$. 

Additional fluctuations from $N_{qp}$ are then needed to explain the measured value, which requires
knowledge of the bulk capacitance $C_L$. The capacitance of the bulk arises from taking the geometric and quantum capacitance in series, $C^{-1}_L = C_g^{-1} + C_q^{-1}$. 
The quantum capacitance can in turn be related to the electronic compressibility of the device, $A \frac{e^2}{C_q} = \frac{\partial \mu}{\partial n}$. 
Fortunately the compressibility of the MLG 1/3-state was measured by our group in Ref.\cite{yang_experimental_2021}, where we obtained $\frac{\partial \mu}{\partial n} \approx  \SI{0.15}{meV \mu m^2}$ at $B = \SI{18}{T}$ (this can be obtained from the published data by extracting the in-gap slope of $\frac{1}{2 \pi \ell_B^2} \frac{\partial \mu}{\partial n}  = \frac{\partial \mu}{\partial \nu} \sim \SI{646}{meV}$).
It remains to convert this to  an estimate of the compressibility at the  $B = \SI{9}{T}$ used here.
Proceeding phenomenologically, it is reasonable to assume the in-gap density of states arises from quasiparticles pinned to impurity sites.  
Assuming the spatial density of these states is set by the impurity density, and is hence independent of $B$,\cite{assouline_energy_2024} with energies equally distributed over the FQH gap $\Delta_{1/3} \propto \sqrt{B}$, we conclude the compressibility will scale as $1 / \sqrt{B}$. We thus scale the compressibility by $\sqrt{9 / 18}$ to obtain $\frac{\partial \mu}{\partial n} \approx  \SI{0.1}{meV \mu m^2}$. The estimated quantum capacitance is then $\frac{e^2}{C_q} \sim \SI{0.13}{meV}$ using the estimated size $A  \sim \SI{0.8}{\mu m^2}$ of the interferometer. The quantum capacitance then adds in series with doubly-gated geometric capacitance $\frac{e^2}{C_g} = \frac{e^2 \epsilon_z d}{2 A} \sim \SI{0.17}{meV}$: all together, we thus estimate $\frac{e^2}{C_L} = \SI{0.3}{meV}$. 

Taking the bulk and edge capacitances in parallel, we obtain $e^2 / C_t \sim (1 / 0.3 + 1/0.234)^{-1} \SI{}{meV} \sim \SI{0.131}{meV}$, giving a dephasing temperature $T_0 = \SI{77}{mK}$, quite close to the measured value of $T_0 = \SI{87}{mK}$. This may be partially a matter of good fortune given the estimates involved in $C_L$, but the scales are clearly consistent with a dephasing temperature set by a combination of bulk and edge contributions.
In the future this analysis could be made quantitative using devices capable of measuring the capacitance of the center gate defining the interferometer. 

\clearpage 
\newpage

\subsection{S5. Phase Slip Magnitude Analysis}

\begin{figure}[hb!]
    \centering
    \includegraphics[width = \textwidth]{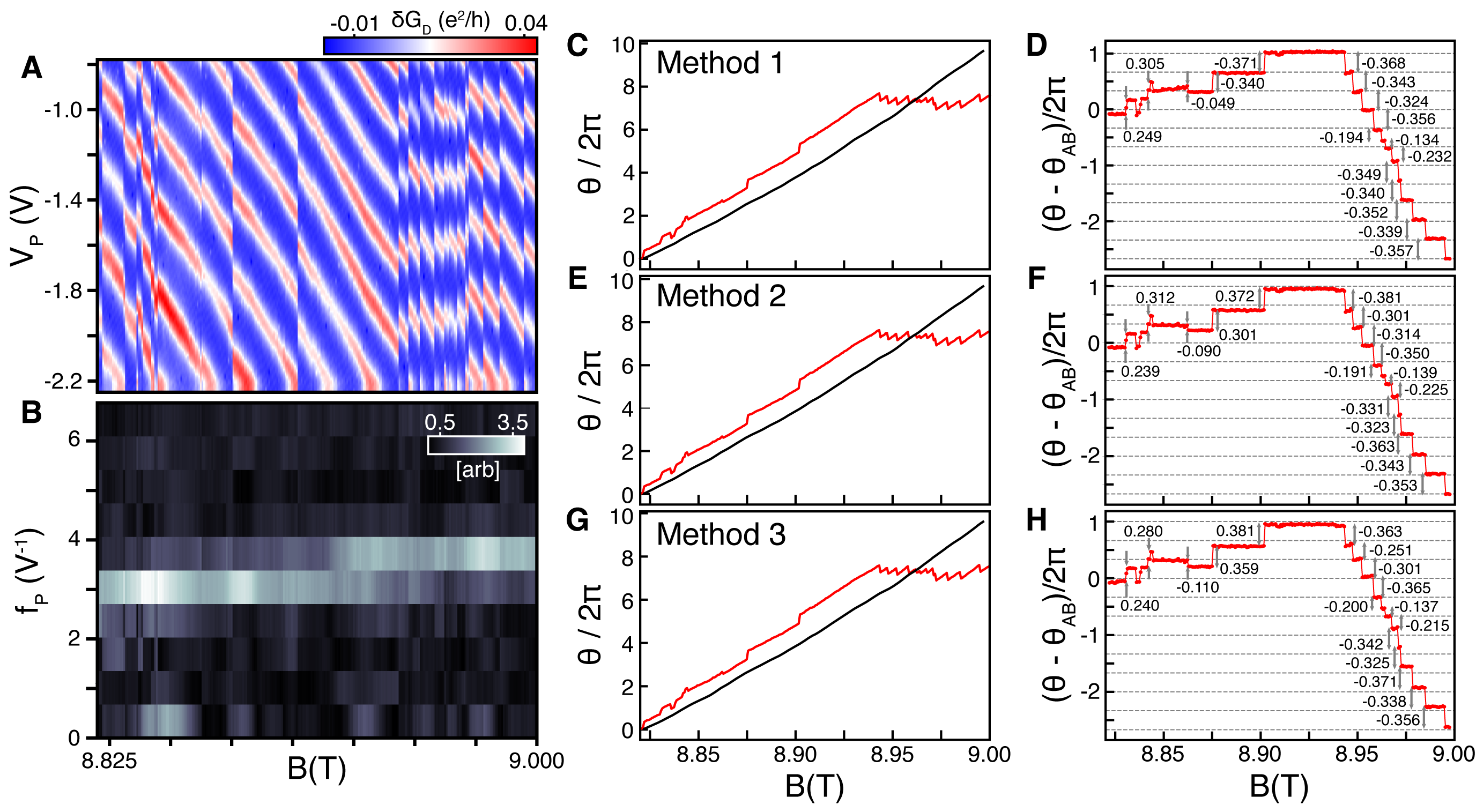}
    \caption{\textbf{Extracting Phase Slip Magnitudes} \textbf{(a)} Same data as in main text Fig.~\ref{fig:fig2}a, but with the mean value of $G_D$ for each value of magnetic field subtracted from the corresponding trace versus $V_P$.  \textbf{(b)} Absolute magnitude of the discrete 1D Fourier transform (along $V_P$) of the data in panel a plotted versus magnetic field and inverse $V_P$.  \textbf{(c, e, g)} The integrated $\theta$ vs. B extracted from the DFT is plotted in red.  The integrated background, which we subtract off from the accumulated phase to calculate the magnitude of the phase slips, is plotted in black.  Panel c shows the integrated phase and the integrated background extracted from the DFT by taking the phase of the largest magnitude component; this is the analysis method we use for all phase slip extractions in the main text.  Panel e shows the same data as panel c, but the phase is extracted by taking a weighted average over the peak in the DFT amplitude.  Panel g shows the same data as panels c and e, but the phase is extracted by taking an unweighted average over the peak in the DFT amplitude.  \textbf{(d)} $\theta - \theta_{AB}$ versus magnetic field extracted by subtracting the integrated background from the integrated phase calculated via the method in panel c.  This data is identical to the main text Fig.~\ref{fig:fig1}e.  \textbf{(f)} Same as panel d, but the phase slip magnitudes are calculated from the data in panel e.  \textbf{(h)} Same as panels f and d, but the phase slip magnitudes are calculated from the data in panel g. 
    }
\label{fig:FFT}
\end{figure}

To extract the magnitudes of the phase slips reported in Fig.~\ref{fig:fig2}b we start by subtracting off the mean value of $G_D$ for each line-trace versus $V_P$ at a fixed value of B.  We then take the 1D discrete Fourier transform (DFT) along the $V_P$ axis of the data in Fig.~\ref{fig:FFT}a (the mean subtracted data from main text Fig.~\ref{fig:fig2}a).  The absolute value of the resulting DFT is shown in Fig.~\ref{fig:FFT}b. We then extract the phase from the component of the DFT which has the largest absolute magnitude, here around 3 cycles per volt of $V_P$.  The point-to-point difference in the phase, extracted in this way, as a function of B contains a nearly constant background, as expected for a linear phase accumulation from the Aharonov-Bohm effect.  However, this constant background is interrupted by sharp spikes at the location of each sudden phase slip.  

To remove the Aharonov-Bohm effect and purely extract the value of the sudden phase slips, we take a moving trimmed mean in a 7-point window (ignoring outliers bigger than $|\Delta \theta| = 0.4$) of the point-to-point difference of the phase versus B.  This enables us to extract a rough estimate of the background.  This rough background is fit to a line, and then further refined with a new trimmed mean of the original point-to-point difference (with the same window size).  The refined trimmed mean excludes outliers greater than the fit value at fixed B plus $\Delta \theta = 0.14$ or less than the fit value minus $\Delta \theta = 0.17$.  The integrated phase (plotted in red) including the sudden slips, along with the integrated background (plotted in black), is shown in Fig.~\ref{fig:FFT}c.  Subtracting the integrated phase from the integrated background produces the dataset shown in Fig.~\ref{fig:FFT}d where the phase jumps are isolated. The outlier exclusion range at the refinement step is set to guarantee the Fourier analysis reproduces the number of observed phase slips in Fig.~\ref{fig:FFT}a.   

While the calculated phase slip magnitudes weakly depend on the choice of the trimmed mean exclusion range, the variation is limited to the standard deviation in the point-to-point phase difference after coarsely removing the phase slips using the same criterion as the initial trimmed mean (outliers with $| \Delta \theta | > 0.4$ are removed).  We find that $\sigma_{\Delta \theta} = 2\pi \cdot 0.012$; leading to an uncertainty of $2\pi \cdot 0.012 \cdot \sqrt{2} = 2\pi \cdot 0.017$ for the magnitude of a sudden phase slip.  While this is small compared to the magnitude of the observed jumps, it is not the dominant source of error in our measurement.  As can be seen by the the resulting DFT in Fig.~\ref{fig:FFT}b, the peak in the DFT amplitude spectrum has some finite width; this is ultimately the result of the interference signal not being perfectly periodic in $V_P$.  Consequently, there is some freedom in the analysis method used to determine the phase of the primary DFT component which introduces some systematic error beyond the statistical contribution.  

To roughly quantify the size of this systematic uncertainty, we chose two additional analysis methods and checked the processed phase slip magnitudes against the method used in the main text.  The second method we used was to take a weighted average of the phase, weighted by the normalized absolute magnitude of the DFT signal, around the peak in the DFT between $f_P = \SI{2}{V^{-1}}$ and $f_P = \SI{5}{V^{-1}}$.  An identical set of panels to Fig.~\ref{fig:FFT}c-d where the phase jumps are extracted using the weighted average (method 2), are shown in Fig.~\ref{fig:FFT}e-f.  The third method we used (method 3) was to take an unweighted average of the DFT phase across the same range in method 2.  Similarly to Fig.~\ref{fig:FFT}c-f, the integrated phase and background, as well as the extracted phase slips, are plotted in Fig.~\ref{fig:FFT}g-h.  We find that for each phase slip, the typical deviation between the three methods is $|\delta (\Delta \theta)| < 2\pi \cdot 0.01$.  However, for some phase slips the deviation can be as large as $ \pm 2\pi \cdot 0.037$ between methods. Since we cannot check every possible analysis method, we will take our global systematic uncertainty to be the largest observed deviation (across all phase slips in Fig.~\ref{fig:FFT}a) from the average across the three methods; we find this to be $\pm 2\pi \cdot 0.04$.  

We use the same method outlined in the beginning of this section (method 1) to extract the sudden phase slip magnitudes for all other relevant datasets. We note that while the statistical uncertainty from the background subtraction is not relevant to other datasets, such as Fig.~\ref{fig:fig5}, \ref{fig:fig6}, and \ref{fig:fig8}, where no Aharonov-Bohm component needs to be subtracted, the systematic uncertainty from the Fourier analysis persists.  Consequently, for datasets where we analyze phase shifts with no background component, we still assume an uncertainty of $\sigma_{\Delta \theta} \approx \pm 2 \pi \cdot 0.04$.

\clearpage
\newpage

\subsection{S6. Estimation of $K_{IL}$ for a compressible puddle}

Here we justify the estimate $K_{IL} \lesssim \frac{1}{2} \frac{e^2}{C^g_{bulk}}$, where $C^g_{bulk} = \frac{\epsilon_z A_I}{d/2}$ is the geometric capacitance of the bulk. The upper bound is expected when the density of added quasiparticles is uniform and approaches right up to the edge of the interferometer; if the puddle is confined inwards, $K_{IL}$ will be decreased exponentially due to screening of the bulk-edge interaction by the gates, as analyzed below.

Recall the couplings $K$ are defined through the phenomenological charging energy $E = \delta n_L^2  K_L/2 + \delta n_L \delta n_I K_{IL} + \delta n_I^2  K_I/2$, where $\delta n_{I/L}$ is the charge added to the edge / bulk. 
$K_{IL}$ is then determined by the following: when charge density $\delta n_L / A_I$ is added to the bulk, what is the resulting potential produced at the edge?
When the radius of curvature is large compared to $d$, we can treat the added charge as an infinite half-plane, with charge distribution  $\delta n(x, y) = \theta(x) e  \delta n_L / A_I$.
Deep in the bulk, $x \gg 0$, this distribution produces potential $\phi = e \delta n_L /  (A_I c_g)$, where $c_g = \epsilon_z / (d / 2)$ is the capacitance per unit area to the double gates. On the other hand, for $x \ll 0$ $\phi = 0$. By reflection symmetry,  $\phi = e \delta n_L / 2 A_I c_g$ at $x = 0$.
For an edge at $x=0$, we thus conclude $K_{IL}  = e \phi(x=0) / \delta n_L \approx e^2 / 2 C^g_{bulk}$.

To determine the correction when the puddle is displaced inwards, we can solve for the full spatial dependence $\phi(x)$.
 We first solve Poisson's equation in the presence of the double gate to conclude a line charge $\rho$ at $x=0$ produces a potential 
\begin{align}
\phi(x) = -\frac{\rho}{2 \pi \sqrt{\epsilon_{xy} \epsilon_z}}\log(\tanh(\pi x \sqrt{\epsilon_{z} / \epsilon_{xy} } / 4 d))
\end{align}
where $d$ is the gate distance and $\epsilon_{xy, z}$ is the anisotropic hBN dielectric constant. 
The total potential produced at the $x=0$ edge by the $x > w$ puddle of density $\delta n_L / A_I$ is thus
\begin{align}
\phi_{IL} =  -e \frac{\delta n_L / A_I }{2 \pi \sqrt{\epsilon_{xy} \epsilon_z}} \int_{w}^\infty dy\log(\tanh(\pi y \sqrt{ \epsilon_z/ \epsilon_{xy}  } / 4 d))
\end{align}

The geometric estimate is thus $K_{IL} = e \phi_{IL} / \delta n_L$, from which  it follows that
\begin{align}
K_{IL} / (e^2 / A_I c_g) &= - \frac{\frac{1}{2 \pi \sqrt{\epsilon_{xy} \epsilon_z}} \int_{w}^\infty dy\log(\tanh(\pi y \sqrt{ \epsilon_z / \epsilon_{xy} } / 4 d))}{\frac{d}{2 \epsilon_z}} \\
& =  - \frac{4 }{ \pi^2} \int_{ \pi \sqrt{ \epsilon_z/\epsilon_{xy}} w/ 4 d}^\infty ds \log(\tanh( s ))
\end{align}
When $w \to 0$, we obtain the limit $K_{IL} / (e^2 / A_I c_g) \to 1/2$. For $w  \gg d$, the bulk edge coupling falls of exponentially due to the screening from the gates, and asymptotically we find
\begin{align}
\lim_{w \gg d} K_{IL} / (e^2 / C^g_{bulk}) \to \frac{4}{\pi^2} e^{-\pi \sqrt{ \epsilon_z/\epsilon_{xy}} w / 2  d}
\end{align}

Assuming e/3 quasiparticles enter the interferometer with at most $\Delta \theta = 0.03$ (the maximum bulk-edge coupling consistent with the phase slip magnitudes observed in Fig.~\ref{fig:fig6}b), and taking $\epsilon_z = 3.25, \epsilon_{xy} = 6.6$, and $ (e^2 / A_I c_g) / K_{IL} = 22.8$, we estimate $w / d =  2.02$, so $w = \SI{90.8}{nm} >> \ell_b$ consistent with the long charging time-scale associated with the observed phase-slips.\\

\clearpage
\newpage 

\end{document}